\documentclass{aa}  
\usepackage[utf8]{inputenc}
\usepackage[english]{babel}
\usepackage{enumitem}
\usepackage{appendix}
\usepackage{gensymb}
\usepackage{multicol}
\usepackage{titlesec}
\titlespacing{\subsection}
  {0pt}    
  {0.8em}  
  {0.4em}  
\titlespacing{\section}
  {0pt}    
  {0.8em}  
  {0.4em}  
\newcommand{\Gaia}{\textit{Gaia}}
\usepackage{graphicx}
\usepackage{txfonts}
\usepackage{threeparttable}
\usepackage{natbib}
\setcitestyle{authoryear, open{((},close{))}}
\newcommand{\citenop}[1]{\citeauthor{#1}, \citeyear{#1}}
\usepackage{rotating}
\usepackage{hyperref}
\hypersetup{
    colorlinks=true,
    linkcolor=blue,
    citecolor=blue,
    }

\begin{document} 

\title{A homogeneous view of asymptotic giant branch carbon stars as seen by \Gaia}

\author{
Alessio Liberatori\inst{1, 3, 4} \and 
Despina Hatzidimitriou\inst{1} \and
Konstantinos Antoniadis\inst{2,1} \and
Giada Pastorelli\inst{3,4} \and
Michele Trabucchi\inst{3} \and
M.~A.~T.~Groenewegen\inst{5} \and
Diego Bossini\inst{4} \and 
L\'eo Girardi\inst{4} \and
Paola Marigo\inst{3}\thanks{Deceased} \and
Alessandro Bressan\inst{4,6,7} \and
Ioannis N. Kallimanis\inst{3,4} \and
Guglielmo Costa\inst{3,4} \and
Vasileios Katsis\inst{8,1} \and
Georgios Vasilopoulos\inst{1} \and
Stamatis Chatzipetros\inst{1}
}

\institute{National and Kapodistrian University of Athens, Panepistimiopolis, 15784 Zografos, Athens, Greece.
\and 
IAASARS, National Observatory of Athens, 15236 Penteli, Greece.
\and
Dipartimento di Fisica e Astronomia Galileo Galilei, Università di Padova, Vicolo dell’Osservatorio 3, I-35122 Padova, Italy
\and
Osservatorio Astronomico di Padova—INAF, Vicolo dell’Osservatorio 5, I-35122 Padova, Italy
\and
Koninklijke Sterrenwacht van Belgi\"e, Ringlaan 3, B-1180 Brussels,
Belgium
\and
SISSA, Via Bonomea 265, I-34136 Trieste, Italy
\and
Purple Mountain Observatory, Chinese Academy of Sciences, Nanjing 210023, People’s Republic of China
\and 
Physique des Interactions Ioniques et Moléculaires, CNRS, Aix Marseille Université, Marseille, France
}

\abstract{\textit{Context. }Carbon stars on the asymptotic giant branch are major contributors to the dust enrichment of galaxies, with gas mass-loss rate values up to $ \dot{M} \approx  10^{-4} \ \rm M_{\odot} \ yr^{-1}$. 
They represent the final evolutionary stage of low- and intermediate-mass stars, during which recurrent dredge-up episodes enrich their atmospheres with carbon and trigger the formation of dust. 
Through their intense winds, they inject large amounts of newly formed carbonaceous dust into the interstellar medium, playing a central role in the chemical evolution of galaxies. 
Their stellar and dust properties have been studied for decades, with a particular focus on the carbon stars in the Magellanic Clouds (MCs).

\textit{Aims. } We aim to homogeneously analyse through the spectral energy distribution (SED) fitting the Gaia DR3 Golden Sample of Carbon Stars, focusing on sources belonging to the Milky Way (MW) and the MCs. 

\textit{Methods. }Our dataset consists of 14,747 stars with complete multi-band photometry from Gaia, 2MASS, and WISE, combined with recent distance and extinction estimates. For a subsample of Mira variables made of 2,494 stars, we also modelled multi-band light curves to obtain accurate mean magnitudes. Stellar and circumstellar properties were derived by fitting the observations with a large grid of synthetic models computed with the DUSTY radiative transfer code, using COMARCS model atmospheres as input. For each target, we determined stellar and dust parameters such as the effective temperature, optical depth, and gas mass-loss rate.

\textit{Results. }The resulting distributions reveal typical effective temperatures around 3150 K. Mass-loss rates range from $\rm 10^{-11}$ to $\rm 10^{-4}$ $\rm M_{\odot}$ $\rm yr^{-1}$. The average dust temperature at the inner bound of the dust shell is about $T_d =1000$ K.
We also observe a correlation between photometric variability amplitude and mass-loss rate.

\textit{Conclusions. }This homogeneous framework provides a unified view of carbon stars across environments spanning a wide range of metallicities, supported by strong statistical coverage. Our results show that some of the physical properties of carbon stars exhibit a dependency on the galactic environment. However, these dependencies do not necessarily reflect intrinsic metallicity effects, but are influenced by differences in luminosity distributions and by the selection biases affecting the available samples. The use of \Gaia\ and WISE introduces combined selection effects that are significant, limiting the detection of both the most dust-enshrouded objects and the less luminous sources in the Magellanic Clouds. While this limits the completeness of the comparison, the observed trends remain statistically robust within the selected samples.}


\maketitle

\section{Introduction}
\label{section:introduction}
Carbon stars are characterized by an atmospheric carbon-to-oxygen number ratio ${\rm C/O}$ larger than 1.
The C/O can either result from extrinsic processes such as the accretion of ${\rm C}$-enriched material, or from physical processes originating in the star itself. 
The latter is the case for Thermally Pulsing Asymptotic Giant Branch (TP-AGB) stars.
During their evolution along the AGB phase, stars with a mass $\rm M\gtrsim1~M_{\odot}$ \citep{di2016studying} experience the third dredge-up. 
This event mixes the products of nuclear burning into the outer envelope, enriching the photosphere with C. 
However, stars with masses $\rm M\gtrsim3.5~M_{\odot}$ are prevented from becoming carbon stars, due to the bottom burning \citep{iben&renzini1983hbb} process that takes place at the base of the convective envelope of these stars and activates the CN cycle that coverts carbon into nitrogen. 
Stars with masses between 1 and 3.5 solar masses \citep{di2016studying} undergo recurrent thermal pulses and dredge-up episodes. 
These mass limits strictly depend on the initial chemical composition and vary according to the adopted stellar evolution models \citep{ventura2016nature, dellagli2017agbdustsolar, pastorelli2019smc, abia2020properties, marigo2022agbopengaia, straniero2023carbonmistery}.
A a series of third dredge up events mix the products of nuclear burning into the outer envelope, enriching the photosphere with C.
At the relatively low temperatures of AGB atmospheres, molecules form efficiently.
Due to the high stability of the carbon monoxide molecule (CO), which consumes the less abundant of the two elements, the C/O abundance ratio becomes the key parameter: C/O $>1$ leads to carbon-rich (C-rich; \citealt{iben1983carbon}) stars, while C/O $<1$ results in oxygen-rich (O-rich) stars. The carbon enrichment increases the opacity of the outer layers, and together with pulsation-driven shocks creates favourable conditions for dust condensation and growth. Radiation pressure on dust grains then drives an intense stellar wind, ejecting gas and dust into the interstellar medium \citep{Hofner2018Review}.

As a result, carbon-rich AGB stars are one of the more important contributors to the enrichment of the interstellar medium (e.g.,~\citenop{marigo2002asymptotic}; \citenop{karakas2007stellar}).
Their dust production is dominated by amorphous carbon (amC), with a minor contribution from silicon carbide (SiC), as confirmed by several studies \citep[e.g.,~][]{groenewegen2022wise}.
Thus, carbon-rich AGB stars play a significant role in the chemical evolution of galaxies and in shaping their observed properties.
Indeed, they account for a substantial fraction of the infrared (IR) luminosity and are key contributors to the integrated light of unresolved galaxies (e.g.,~\citenop{maraston2006evidence}; \citenop{marigo2009tp}; \citenop{boyer2011surveying}; \citenop{melbourne2013contribution}). 
In this work, we focus on intrinsic carbon-rich AGB stars (resulting from the third dredge-up) and, for the sake of brevity, we will hereafter refer to them simply as 'carbon stars' (or C-stars).

The properties of carbon stars make them promising distance indicators. In particular, the J-region asymptotic giant branch (JAGB) method \citep{freedman2020astrophysical} uses a subset of these stars as standard candles. This approach has been shown to provide distance estimates consistent with those from the tip of the red giant branch (TRGB) method \citep{lee1993tip}, with recent applications confirming its accuracy \citep{lee2023carbon}. Moreover, thanks to their higher luminosity, JAGB stars can be detected at greater distances than TRGB stars \citep{lee2024resolved}, offering a potential extension of the extragalactic distance scale and contributing to the calibration of the Hubble constant.

The advent of large-scale surveys such as \Gaia\ \citep{gaia2016gaia}, the Two Micron All Sky Survey (2MASS) \citep{skrutskie2006two}, and the All-Sky Release Catalog \citep[AllWISE, ][]{cutri2021vizier} of the Wide-field Infrared Survey Explorer \citep[WISE, ][]{Writhg_etal_2010}, has enabled the identification and characterization of a large number of carbon stars across the Milky Way (MW) and in the Magellanic Clouds (MCs). Taken together, these surveys provide extensive and homogeneous photometric coverage, from 0.5 to 22 microns, allowing a more reliable determination of the properties of carbon stars across different stellar populations and galactic environments.

In the last decade, numerous studies have been conducted to determine the properties of carbon stars by using the spectral energy distribution (SED) fitting method through multi-band photometry data (e.g.,~\citenop{Bressan1998mod}; \citenop{groenewegen2009luminosities}; \citenop{riebel2012mass}; \citenop{boyer2013there}; \citenop{srinivasan2016identifying};\citenop{rau2017adventure}; \citenop{nanni2018estimating}; \citenop{groenewegen2018luminosities}; \citenop{nanni2019mass}; \citenop{nanni2019optical}; \citenop{groenewegen2022wise}).
By comparing the observed photometric data to the available synthetic models, these studies have been able to constrain key quantities such as mass-loss rates, dust properties, luminosities, and effective temperatures of stars using multi-band photometry in optical and IR bands.

A large variety of synthetic models are available for this purpose in the literature.
Among the hydrostatic stellar atmosphere models used to describe stellar spectra are COMARCS \citep[built specifically for AGB C-rich or O-rich stars,~][]{aringer2016synthetic, aringer2019carbon} or MARCS \citep{gustafsson2008grid}. 
These models are subsequently combined with radiative transfer codes that incorporate dust features into the stellar spectral energy distribution (SED). 
Commonly used softwares include DUSTY \citep{ivezic1997self}, often applied together with COMARCS models \citep[e.g.,~][]{nanni2018estimating, nanni2019mass, groenewegen2022wise}, and 2Dust \citep{ueta20032dust}, which provides the basis for the GRAMS models \citep{sargent2011mass, Srinivasan2011massloss} constructed from MARCS and COMARCS atmospheres \citep[e.g.,~][]{riebel2012mass, boyer2013there}.

The great majority of the aforementioned studies investigated the properties of carbon stars in the Magellanic Clouds (e.g,~\citenop{groenewegen2009luminosities}, \citenop{riebel2012mass}, \citenop{rau2017adventure}).
One reason is due to the star formation history characterizing these two galaxies, which has produced numerous stars now observed as carbon stars stars (see \citenop{blanco1983distribution} for a detailed study of carbon stars as tracers of past star formation activity in the LMC).
At the same time, the interpretation of the data is facilitated by the known distances and extinction, which are often much less well constrained within the Milky Way.
Indeed, only a limited number of studies concentrate on determining stellar properties for the MW, with notable examples being \cite{nanni2019optical} and \cite{groenewegen2022wise}.

With the advent of \Gaia, obtaining highly accurate distance measurements for Galactic stars has become possible, enabling more reliable extinction estimations and more accurate determinations of stellar parameters.

In this study, we analyze the \Gaia\ DR3 Golden Sample of Carbon Stars \citep{Creevey2023gaiagoldensample}, deriving stellar and dust properties for its sources.
This dataset includes carbon stars in the Milky Way and in the Magellanic Clouds, and allows for a consistent treatment of stars formed in different galactic environments and with different metallicities. This approach enables us to investigate populations that differ in metallicity and evolutionary history within a single, self-consistent framework, ensuring that results obtained for the different systems can be directly compared without biases introduced by heterogeneous selection or analysis methods. In particular, we place special emphasis on the Galactic sample, for which the availability of precise \Gaia\ parallaxes and recent extinction estimates allows a reliable determination of stellar and dust parameters. 
Carbon stars can also be Mira variables, for which variability can strongly affect the construction of spectral energy distributions. To account for this, we modeled multi-band light curves for a subset of Miras, and we used the mean magnitudes derived from the best-fit models in place of single-epoch \Gaia\ photometry, ensuring consistent and variability-corrected inputs for the analysis.

This paper is organized as follows. Section~\ref{section:dataset} describes the dataset and the photometric sources selected for this work. In Section~\ref{section:lightcurves}, we detail the method applied to correct the photometry for variability effects in Mira variables. Section~\ref{section:models} introduces the grid of models adopted for the SED fitting, while Section~\ref{section:Analysis} outlines the fitting procedure itself. We report our findings in Section~\ref{section:results} and compare them with the literature in Section~\ref{section:discussion}, addressing the relevant caveats and physical implications. Finally, our conclusions are summarized in Section~\ref{section:conclusion}.

\section{Dataset}
\label{section:dataset}
Several catalogues of carbon stars have been compiled and published over the past thirty years, many of which are based on a combination of earlier studies. This has resulted in a complex landscape of partially overlapping datasets. 
Among the most recent efforts is the work by \cite{suh2024infrared}, which integrated multiple catalogues with stricter selection criteria to construct a refined sample of AGB carbon stars.
Additional examples of recent C star catalogues are, for instance, \cite{abia2022characterisation}, \cite{chen2012carbon}, and \cite{li2018carbon}.

Despite efforts to standardize selection criteria, many of the catalogues discussed above still suffer from limitations in homogeneity. 
This study emphasizes homogeneity as a fundamental criterion by adopting the Gaia DR3 Golden Sample of Carbon Stars (GGSCS, hereafter) as a sourcelist, for which we infer both stellar and dust properties.
Besides being homogeneous, the GGSCS has extensive sky coverage, and high astrometric precision. 
The authors of this catalogue identified the stars based on the strength of C$_2$ and CN molecular bands in their \Gaia\ XP spectra. 
Stars with colours bluer than \( G_{BP} - G_{RP} = 2 \) or fainter than \( G = 17.65 \rm~mag\) were excluded as detailed by the \citet{Creevey2023gaiagoldensample}. 
The catalogue includes a total 15,740 sources in the Milky Way and the Magellanic Clouds.

\begin{figure}[!h]
    \centering
    \includegraphics[scale=0.54]{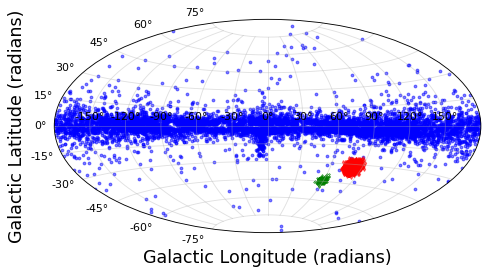}
    \includegraphics[scale=0.5]{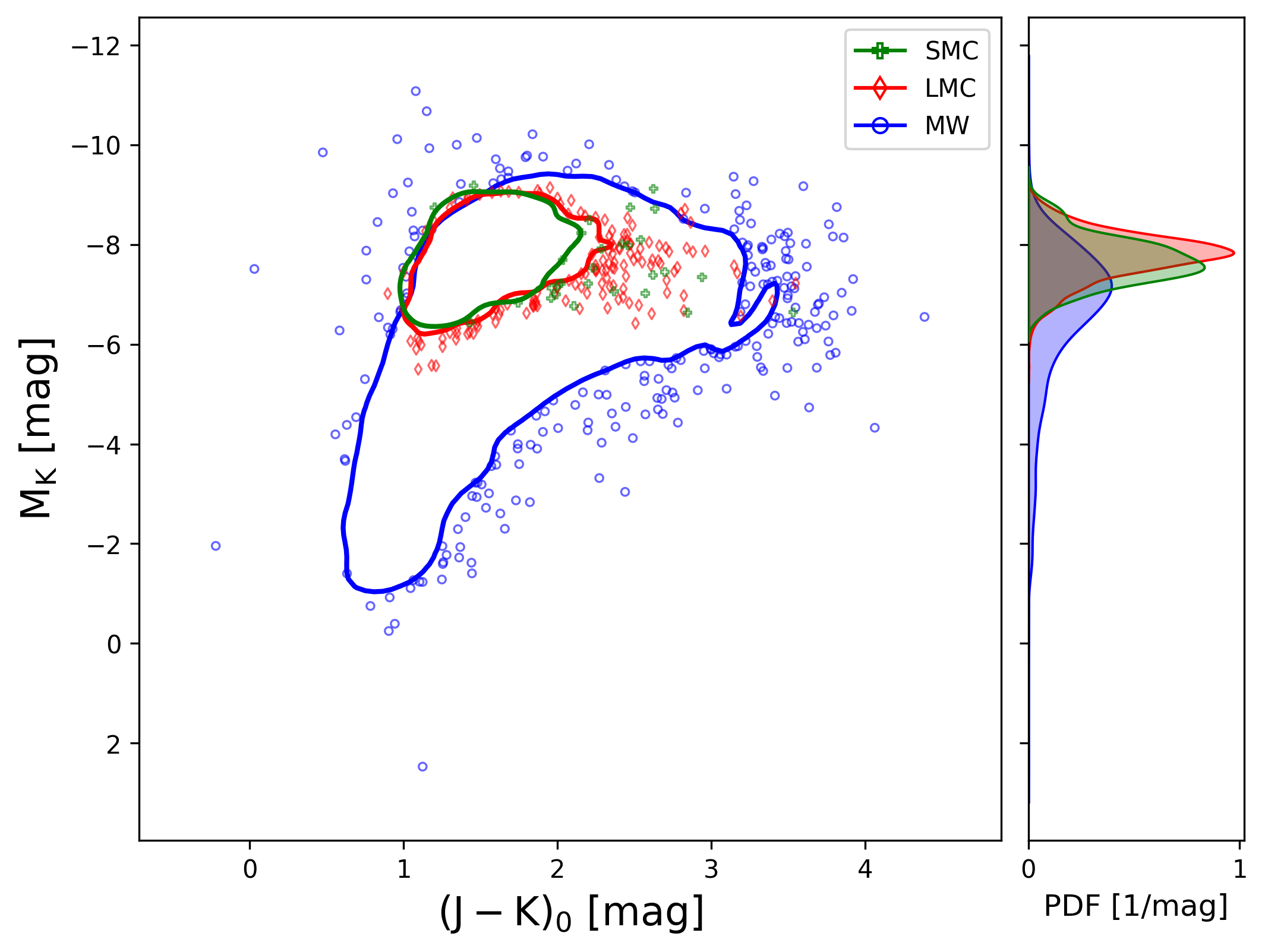}
    \caption{Top panel: Sky distribution of the targets in the final catalogue. Bottom panel: Extinction-corrected color-absolute magnitude diagram ($J-K_s$ vs. $M_{K_s}$). The right sub-panel shows the histograms of the absolute $K_s$ magnitude probability density function (PDF) for the three populations. The color coding is as follows: SMC stars in green, LMC in red, and MW in blue.}
    \label{fig:sky-cmd}
\end{figure}

Figure~\ref{fig:sky-cmd} shows the sky distribution of the sources and their location on a JHK Color-Magnitude Diagram (CMD).

\subsection{Catalogue assembly}
We complemented \Gaia\ photometry with infrared data by matching the positions of the sources in the GGSCS with the 2MASS and AllWISE catalogues using a 1.0-arcsec search radius via the CDS Xmatch Service\footnote{\url{http://cdsxmatch.u-strasbg.fr/}}. We find 15,096 sources having a counterpart in both catalogues, and we retain 15,072 stars having all the 10 filters values available.

The three \Gaia\ photometric filters are sensitive to stellar effective temperature ($T_{\rm eff}$) and dust absorption, while the WISE bands are sensitive to dust emission and can trace chemical signatures, such as the silicon carbide feature at 11.3~$\rm \mu m$ in the W3 band, in the circumstellar envelopes. 
The 2MASS bands are sensitive to both the stellar $T_{\rm eff}$ and to dust emission.
We note that \Gaia\ and WISE magnitudes partially account for photometric variability through multi-epoch averaging, whereas this is not the case for 2MASS observations that may not accurately represent the average magnitude of variable sources.
Section~\ref{section:lightcurves} discusses this aspect.

A number of studies similar to ours \cite[e.g.,~][]{riebel2012mass, groenewegen2018luminosities} also employ photometric data from \textit{Spitzer} \citep[filters I1, I2, I3, I4]{werner2004spitzer}. We preferred not to use these data in order not to compromise the homogeneity of our sample, as only 41\% of the stars have at least one \textit{Spitzer} filter available.

We verified our positional crossmatch against the catalogue by \citet{gavras2023crossmatch}, which uses a more advanced algorithm incorporating observational and astrophysical parameters beyond positional coincidence. A total of 12,679 out of 15,072 (84\%) stars in our sample share identical counterparts with those reported by \citet{gavras2023crossmatch}, confirming the reliability of our associations. 
The remaining sources are primarily located in high-density regions.
Our final catalogue explicitly flags the sources found by \citet{gavras2023crossmatch}.

\subsection{Distances and extinction}
Distances are a key parameter in our analysis, as they play a crucial role in both the extinction estimation and the SED fitting (see Section~\ref{section:Analysis}). 
We adopt the distances from \citet{bailer2021estimating} for stars not located in the MCs, acknowledging the inherent limitations of this approach, such as the adoption of Galactic priors and the single-star assumption, as discussed in detail in the original work. For stars in the LMC, we adopted a distance of 49.97 kpc \citep{pietrzynski2013eclipsing}, while for the SMC we used 62.44 kpc \citep{graczyk2020distance}.

For all stars in our catalogue, we estimated integrated extinction values using the 3D dust extinction map of \cite{lallement2022dustmap} combined with distance estimates from \cite{bailer2021estimating}. Unlike asymptotic extinction, this approach traces the dust density along each line of sight, yielding a a more accurate determination of the foreground extinction, $A_V$, rather than relying on the total asymptotic values. The resulting $A_V$ is then converted to other photometric bands (e.g., $A_{BP}$, $A_{K}$, $A_{W1}$) using the tabulated extinction coefficients provided by \cite{wang2019optical}. These authors caution that these extinction coefficients were derived from red clump stars, and may not be appropriate for stars with a significantly different spectral energy distribution. We compared the extinction we derived with the values given by \cite{abia2022characterisation}, and found them to be compatible, suggesting that systematic uncertainties should be small enough not to impact the results and conclusions of the present analysis.
The same extinction prescription was also adopted for the Magellanic Cloud sources. Although dedicated reddening maps for the Clouds are available (e.g.,~\citealt{skowron2021}), we verified that the mean extinction values derived from the \cite{lallement2022dustmap} map along the corresponding lines of sight are fully consistent with those reported by \cite{skowron2021} for both the MCs. Therefore, the use of a homogeneous extinction treatment for all sources does not introduce significant systematic differences in our analysis.

For Galactic sources with larger inferred distances and the sources in the MCs, we therefore adopt the maximum integrated extinction along the line of sight provided by the map. As a consequence, the extinction for these objects may be slightly underestimated; however, this effect is expected to be small and does not qualitatively affect the results presented in this work.

\subsection{Membership}
To determine the membership of stars in the MCs, we applied a clustering algorithm to \Gaia\ DR3 coordinates and proper-motions. To make it more resilient against outliers, use the RobustScaler approach for scaling the data, following a similar application to open clusters by \cite{hunt23members}. We then applied the HDBSCAN algorithm \citep{dbscan1996}, identifying 5,235 LMC stars and 979 SMC stars. A total of 326 stars were found that do not belong to the MCs, but lack \cite{bailer2021estimating} distance estimates.
Thus, we excluded these stars from our analysis. We also excluded 2 sources lacking extinction estimates. Our final sample consists of 14,747 sources, each with complete photometric coverage across all \Gaia, 2MASS, and AllWISE filters, as well as reliable distance and extinction estimates.

\subsection{Selection biases}
\label{subsec:selection_biases}
The construction of the sample relies on \Gaia\ detections and therefore unavoidably introduces selection effects.
Due to its optical sensitivity, \Gaia\ is biased against carbon stars affected by strong circumstellar extinction, such as sources undergoing very high mass-loss rates and surrounded by thick dust shells, which may fall below the \Gaia\ detection threshold.
At the same time, intrinsically faint carbon stars with low mass-loss rates may also remain undetected, particularly at the distance of the Magellanic Clouds, even in the absence of significant circumstellar extinction.
The impact of these distance-dependent selection effects is clearly illustrated in Fig.~\ref{fig:sky-cmd}. 
The three samples cover different ranges of of K-band absolute magnitude, with the SMC distribution showing a more pronounced truncation at the faint end compared to the MW, confirming that Gaia primarily detects only the brightest sources at larger distances.
As a result, the sample may be incomplete at both the low and high ends of the mass-loss distribution, with these effects being substantially more severe in the Magellanic Clouds than in the Milky Way.

Additional selection effects are introduced by the use of WISE photometry.
While WISE is essential for identifying dusty AGB stars, it is subject to sensitivity limits, source confusion, and saturation, especially at longer wavelengths and in crowded regions.
These effects can affect both the faintest sources and the brightest objects with extreme infrared emission.

Consequently, the final catalogue should be regarded as representative of the carbon-star population within the dynamic range jointly accessible to optical and mid-infrared surveys.
Because these selection effects differ between the Milky Way and the Magellanic Clouds, it is not straightforward to directly compare the samples.
Therefore, to properly investigate metallicity-driven effects, it is necessary to account for differences in luminosity and evolutionary stage. 
We plan to address this specific analysis in a follow-up paper.

\section{Average magnitude correction for Mira variables}
\label{section:lightcurves}
Carbon stars are long-period variables (LPVs), often identified as Miras. For such objects, single-epoch photometry may not provide a reliable estimate of the mean magnitude, potentially introducing systematic biases in the derived stellar parameters.

To account for the variability of these stars, we analysed light curves from two main sources: Gaia DR3 epoch photometry (in the $G$, $BP$, and $RP$ bands) and the OGLE III \citep{soszynski2009lpvmagellanic} and OGLE-IV \citep{iwanek2022miramilkyway} survey (in the $V$ and $I$ bands)\footnote{Data have been retrieved from \url{https://ftp.astrouw.edu.pl/ogle/}}. Although \Gaia\ DR3 provides mean magnitudes based on multi-epoch observations, these are not derived from dedicated light curve fitting and may still be affected by variability-related biases. The OGLE data provide densely sampled optical light curves, making them particularly well suited for detailed variability analyses.

For all stars, we performed multi-band light curve fitting using a custom Python code developed for this work. The procedure builds on the \texttt{LombScargleMultiband} implementation \citep{vanderplaslombscarglemultibnd}, combined with additional publicly available tools for multi-band variability analysis. It simultaneously fits all available filters for each star, ensuring a consistent period, phase, and amplitude across bands. The resulting models provide an empirical, self-consistent description of the variability as a function of time.
From the best-fit light curve models, we derived the mean magnitudes in the three \Gaia\ bands ($G$, $BP$, and $RP$). These mean magnitudes are then adopted as input for the SED fitting procedure, described in Section~\ref{section:Analysis}. 
By construction, they are more robust than simple averages of individual epoch measurements, since they incorporate the full variability pattern across filters. 
For stars with OGLE data, we also used the $V$ and $I$ mean magnitudes derived from the same procedure to increase the reliability of the \Gaia\ light curves fit.
In the fitting procedure, we did not take into account \Gaia\ photometric data with the columns \texttt{rejected\textunderscore by\textunderscore photometry} and \texttt{variability\textunderscore flag\textunderscore g\textunderscore reject} as true.

We applied light curve corrections to a subset of stars classified as Miras, representing 18\% of the total sample. These stars are regular pulsators, and the light curve fit is more reliable for this type of star than for the non-regulars.
The identification of Miras was carried out by combining information from the OGLE catalogue with our Gaia-based analysis. For stars included in both Gaia and OGLE, we directly adopted the OGLE classification of Mira-type variables. To extend this classification to stars with only Gaia data, we examined the distribution of OGLE Miras in the diagram of period (P) versus variability amplitude in RP ($Amp{RP}$), finding that most occupy a region defined by an $Amp{RP} > 0.2$~mag and by $100 < P < 650$ days. These empirical boundaries were then applied to the Gaia-only sources to identify additional Mira candidates. The final Mira subsample thus comprises both OGLE-flagged Miras and \Gaia-only stars selected according to these criteria.

The periods derived from our light curve fits are fully consistent with OGLE values, confirming the reliability of our selection.
Appendix~\ref{appendix:lightcurve} discusses in more detail the technical aspects of the algorithm adopted, as well as some light curve fitting examples.
\\

\begin{figure*}[!h]
    \centering
    \includegraphics[scale=0.4]{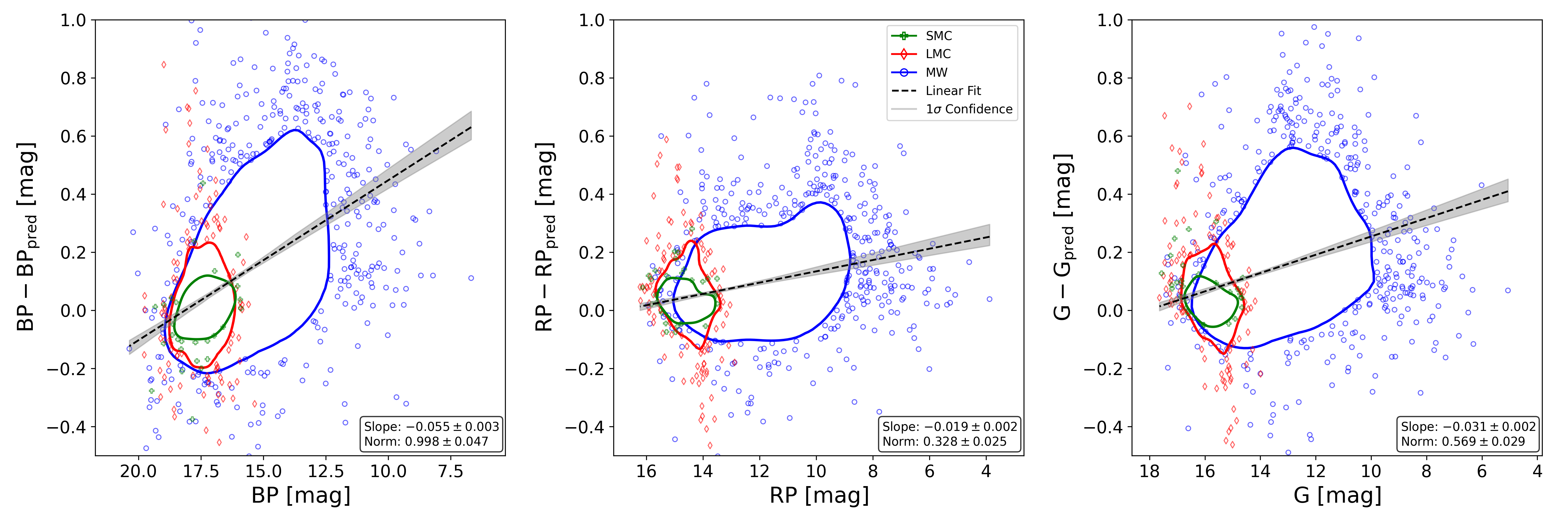}
    \caption{Absolute difference between the un-reddened \Gaia\ BP, RP and G magnitudes and our mean magnitudes derived from the light curve fit. Colours are the same as in Fig.~\ref{fig:sky-cmd}. The y-axis range is restricted to highlight the region of interest, excluding a few outliers. The dashed black line shows the linear fit, and the shaded region denotes the 1$\rm \sigma$ confidence interval of the regression. The resulting fit parameters (slope and intercept) and their uncertainties are reported within each panel.}
    \label{fig:hist_lc_correction}
\end{figure*}

To illustrate the effect of the correction, Fig.~\ref{fig:hist_lc_correction} compares the mean magnitudes provided by \Gaia\ DR3 \citep{GaiaDR3} to the values obtained from our multi-band light curve fitting. 
The left panel of Fig.~\ref{fig:hist_lc_correction} (BP band) exhibits a clear tilt, indicating that the applied correction increases for brighter BP magnitudes, while the effect is much weaker in RP and intermediate in the $G$ band.
This behaviour may reflect the strong optical variability of Mira variables, which is more pronounced in the BP band than in RP, combined with the fact that \Gaia\ magnitudes are flux averages and do not account for the light curve shape.
At fainter magnitudes, increased photometric noise and sparser effective sampling tend to wash out this effect, reducing the apparent correction.

The correction is generally small, with absolute differences below 0.2 mag for most stars, although a few hundred objects show larger corrections, occasionally exceeding 1 mag. 
Corrections exceeding 1 mag occur only in the $BP$ band (and partly in $G$), while they are much smaller in $RP$ and negligible in the near-infrared.
The total number of Miras identified is 2494.
While these cases represent a small fraction of the sample, applying the correction ensures greater reliability of the SED fitting—both statistically, by improving the overall homogeneity of the dataset, and individually, by refining the estimates for Mira-type variables. 
The results of the light curve fits, such as the magnitude and the period found, are available in electronic format at the CDS.

\section{Methods}
\label{section:models}
In order to characterize the catalogue, we constructed a grid of spectral energy distribution models representative of carbon stars, and performed a fit them to the photometric data. In the present section, we describe first the models computation, and then the fitting method.

\subsection{Model grid}
\label{subsection:grid}
In general, the observed SED of carbon stars results from the photospheric stellar emission being reprocessed by circumstellar dust. To model this, we adopt synthetic stellar spectra from the COMARCS\footnote{For further information regarding the model specifics, we refer to the COMARCS website \url{http://stev.oapd.inaf.it/atm//}} library \citep{aringer2016synthetic,aringer2019carbon}, that provides hydrostatic stellar atmospheres specifically developed to study the structure and spectral properties of AGB stars. It is based on the MARCS code \citep{gustafsson2008marcsmodels} and includes updated molecular opacities and line lists, including for C-rich compositions. In this work, we employed all COMARCS models that satisfy the following criteria: $\mathrm{M < 5~M_{\odot}}$, C/O~$\geq$~1, $T_\mathrm{eff}$~$\le$~5000~K, and $T_\mathrm{eff}$~$\geq$~2500~K.

We used COMARCS photospheric spectra as input for DUSTY (V4)\footnote{\url{https://github.com/ivezic/dusty}} \citep{ivezic1999dusty}, a 1D radiative transfer code for modelling the radiation from astrophysical sources reprocessed by a dust shell. It solves the radiative transfer problem for spherical geometry using scaling properties that minimize free parameters, with dust temperatures determined from local radiative equilibrium. In addition to the photospheric spectrum, DUSTY requires information on the dust density structure to predict the output SED. In particular, it can either be computed self-consistently by solving the wind hydrodynamic equations (mode \texttt{RDW}), or an analytical profile can be specified to approximate the density profile. In principle, the former is more appropriate for AGB stars \citep{elitzur2001dusty,nikutta2014wise,Hofner2018Review}, but it is considerably time-consuming. Therefore, the latter is often adopted in the literature \citep[e.g.,~][]{groenewegen2018luminosities} due to its computational efficiency. In order to consistently compare our results with the literature, we also adopted the analytical profile, namely a $r^{-2}$ power law.

For similar reasons, we describe the composition and optical properties of the dust in accordance with the recent literature \citep[e.g.,~][]{groenewegen2009luminosities,groenewegen2018luminosities,riebel2012mass}. 
We adopt a single chemical mixture made of amorphous carbon (amC). 
The optical constants are taken from the ACAR sample provided by \citet{Zubko1996}.
For simplicity, we assumed a single sublimation temperature $T_{\rm sub}=1200$~K. 
The grain size is assumed being 0.1~$\rm \mu m$.
For each COMARCS spectrum, we compute multiple DUSTY SEDs by varying the dust properties, namely the dust temperature and optical depth $\tau_V$ of the dust envelope. 
While the sublimation temperature of amorphous carbon is fixed at 1200~K, the inner dust temperature ($T_{dust}$) is an input parameter that defines the temperature at the inner radius of the dust shell.
The latter, in particular, is a key parameter for determining the influence of dust on the observed properties of the stars. Its value correlates with the amount of dust around the star, thus with the IR excess of the light coming from the SEDs of the stars. Following \citet[][and references therein]{groenewegen2022wise} we assume a fiducial wavelength of 0.55~$\mu{\rm m}$ for $\tau_V$. 
We computed models for five different dust temperatures, from $T_{\rm dust}=400$~K to $1200$~K in steps of $200$~K, and for 55 values of the optical depth of the dust, ranging from $10^{-4}$ to $10$. These values, together with the properties of the adopted COMARCS spectra, define the grid nodes, that are summarized in Table~\ref{tab:model_grid}, for a total of $\sim$ 51,000 models.

The mass-loss rate $\dot{M}$ is computed using the equation from \cite{gullieuszik2012vmc} (originally described by \citealt{groen1998carbonmiras}):
\begin{equation}
\centering
    \tau_{V} = 5.405\times 10^{8} \, \frac{\dot{M} \, r_{gd} \, Q_{V}/a}{r_d \, R_{\star} \, v_{\mathrm{exp}} \, \rho_d}
    \label{eq:massloss}
\end{equation}
where $\dot{M}$ is the mass-loss rate in $\mathrm{M_{\odot}\,yr^{-1}}$, $v_{\mathrm{exp}}$ is the shell expansion velocity in $\mathrm{km\,s^{-1}}$ , and $R_{\star}$ is the stellar radius in solar radii ($R_{\odot}$). Furthermore, $\rho_d$ is the grain density in $\mathrm{g\,cm^{-3}}$, $Q_V$ is the absorption efficiency factor in the $V$ band, $a$ is the dust grain radius in $\mathrm{cm}$, $r_{gd}$ is the dust-to-gas mass ratio, and $r_d$ is the distance to the dust shell in stellar radii.
For the calculation, we fixed $v_{exp} = 10~\,\mathrm{km\,s^{-1}}$ \citep[e.g.,~][]{groenewegen2009luminosities, groenewegen2018luminosities}, we calculated $Q_V$ using the python package \texttt{miepython}\footnote{miepython: Pure python calculation of Mie scattering, \url{https://doi.org/10.5281/zenodo.7949403}}, and we assumed a $\rm \rho_d = 2.2 ~ g ~ cm^{-3}$.
We derived the optical depth $\tau_V$ from the best-fit DUSTY model.
We assumed a constant dust-to-gas ratio of $r_{\mathrm{gd}} = 1/200$ for the MW, LMC, and SMC, consistent with the standard approach adopted in the majority of comparable studies in the literature.
The inner radius $R_{in}$ is derived from the best-fit DUSTY model and subsequently rescaled accordingly to the stellar luminosity, since DUSTY computes this quantity assuming a reference luminosity of $10^4 \mathrm{L_{\odot}}$. DUSTY outputs the inner radius of the dust shell in centimetres, which corresponds to the product of the stellar radius $R_{\star}$ and the dimensionless parameter $r_d$. 
Thus, in Eq. \ref{eq:massloss}, we calculate $R_{\star}r_d=R_{in}=R_{DUSTY}\left(\frac{L_{\star}}{10^4 L{\odot}} \right)^{1/2}$.
The luminosity of each star has been calculated by integrating its SED, and then multiplying it by $\rm 4\pi d^2$.

The synthetic grid comprises over 50,000 models, with input parameters summarized in Table \ref{tab:model_grid}.
\begin{table}
    \centering
    \small
    \caption{Parameters of the synthetic model grid used in this work.}
    \label{tab:model_grid}
    \begin{tabular}{c c}
        \hline\hline 
        Parameter & Range / Values \\
        \hline
        \multicolumn{2}{c}{COMARCS parameters} \\
        $T_{\mathrm{eff}}$~[K] & 2500 -- 5000\tablefootmark{a} \\
        Stellar mass [$\rm M_\odot$] & $M < 5 M\odot$ \\
        C/O ratio & $> 1$ \\
        \hline
        \multicolumn{2}{c}{DUSTY parameters} \\
        Dust type & amC (100\%) \\
        Optical depth $\tau_{0.55\mu m}$ & 0.0001 -- 10 (55 steps)\tablefootmark{b} \\
        Dust temperature [K] & 400 -- 1200 (steps of 200~K) \\
        \hline
    \end{tabular}
    \tablefoot{
        \tablefoottext{a}{Approximate range, irregularly spaced.}
        \tablefoottext{b}{Logarithmic spacing between 0.0001 and 0.1; linear spacing up to 10.}
        
    }
\end{table}

\subsection{SED Fitting}
\label{section:Analysis}
The stellar and circumstellar parameters of the sources in our sample were derived by fitting the observed photometric data to synthetic model SEDs, with the best-fit model for each star determined by minimizing the reduced $\chi^2$ value. The fitting procedure follows the classical $\chi^2$ formalism:
\begin{equation}
\label{equation:chisquared}
\centering
\chi^2_{\rm red} = \frac{1}{N - p} \sum \frac{\left[ \log F(\text{Obs}, \lambda) - \log F(\text{Model}, \lambda) \right]^2}{\left( \frac{\sigma_{F(\text{Obs}, \lambda)}}{F(\text{Obs}, \lambda) \ln 10} \right)^2} ,
\end{equation}
where, $F(\lambda)$ is the flux at a specific wavelength, $N$ is the number of photometric data points for each source, $p$ is the number of free parameters, and $\sigma$ represents the uncertainty on that measurement.
Synthetic fluxes are obtained by convolving the model spectra with the transmission profiles of each observational filter\footnote{The filter profiles were retrieved from \url{http://svo2.cab.inta-csic.es/theory/fps/}.}.

To estimate uncertainties for the derived properties, we stochastically resampled the photometric data and repeated the SED fit 150 times for each star. For each star and filter, the observed magnitude was modeled as a Gaussian distribution centered on the measured value with a 1$\sigma$ width equal to the photometric error reported in the corresponding reference catalogue. 
Random draws from these distributions were used to perform $\chi^2$ fits, yielding a distribution of stellar parameters.

\begin{figure*}
    \centering
    \includegraphics[width=0.3\linewidth]{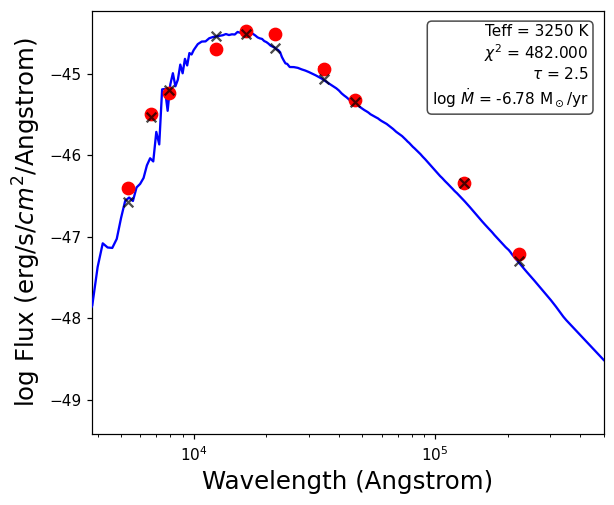}
    \includegraphics[width=0.3\linewidth]{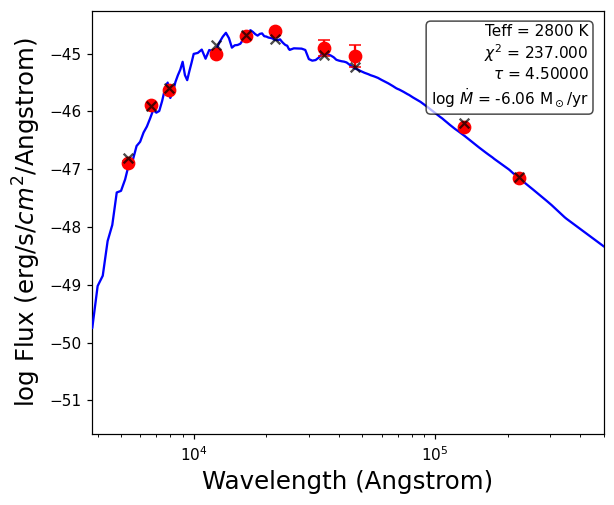}
    \includegraphics[width=0.3\linewidth]{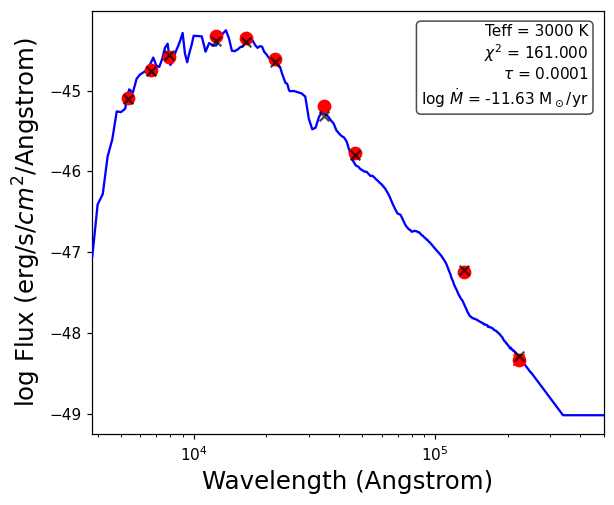}
    \includegraphics[width=0.3\linewidth]{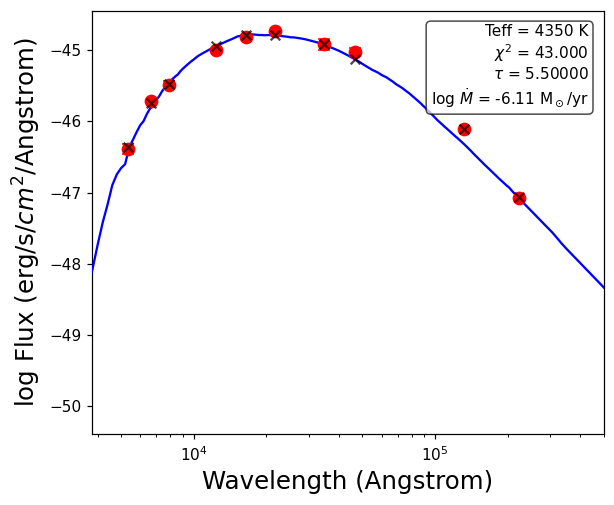}
    \includegraphics[width=0.3\linewidth]{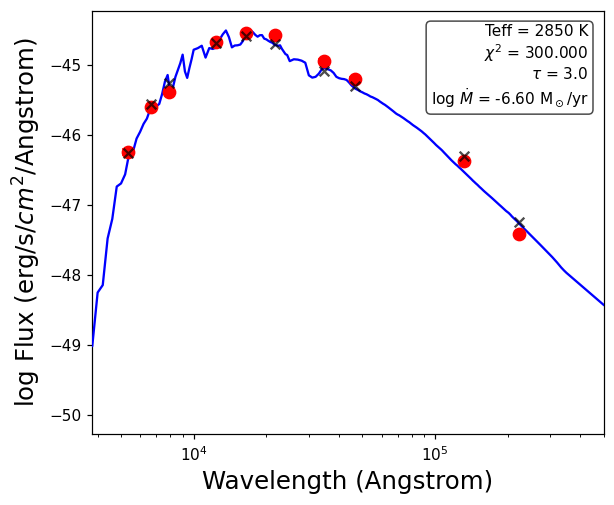}
    \includegraphics[width=0.3\linewidth]{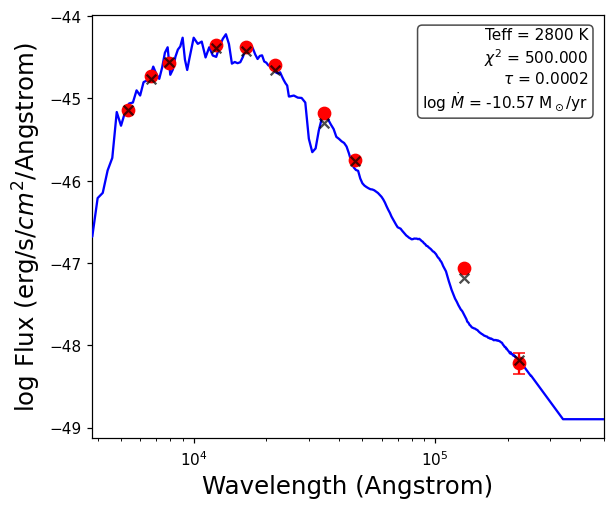}
    \caption{Examples of SED fits for six stars with different stellar and dust parameters. Red points indicate the observed fluxes normalized to the stellar luminosity, the blue line shows the best-fitting model, and the small black crosses represent the model fluxes convolved with the observational filter transmission curves, which are the quantities used in the SED fitting.}
    \label{fig:twoseds}
\end{figure*}

The final parameter values correspond to the medians of these posterior distributions, while the median-absolute-deviation times 1.4826 (one sigma in a Gaussian distribution) quantify their uncertainties. Using the median minimizes the impact of outliers, and we verified that median and mean values closely agree. This approach provides robust parameter estimates and a reliable assessment of their uncertainties.
Figure~\ref{fig:twoseds} shows six examples of fitted SEDs.

To assess the goodness of fit, we considered both the minimum $\chi^2$ value and the average offset between the observed photometric data and the best-fitting model. 
Approximately 95\% of the stars have $\chi^2 < 2000$; only these are retained for our analysis and will be included in the published dataset.

We note that the absolute value of the reduced $\chi^2$ in SED fitting cannot be interpreted in the classical statistical sense, as it strongly depends on the number of photometric points, on the adopted uncertainties, on the model grid resolution, and on the intrinsic variability of AGB stars.
For this reason, $\chi^2$ values from different studies cannot be directly compared.
In our work, $\chi^2$ is used exclusively as a relative internal metric to rank the goodness of fit among models for the same source.

Assuming that a perfect agreement between observed and model fluxes (i.e. an average flux ratio of unity) gives a ratio equal to one, we find that 97\% of the stars show an average ratio $\rm 0.9955<\overline{\left(\frac{Obs}{Model}\right)}<1.0045 ~ erg/s ~ cm^2$, indicating generally good fits across the sample.
At the end of the chi-squared process, we find a best-fit model for 13,777 sources (93 \% of the total). These stars compose the dataset that we discuss and publish.

\section{Results}
\label{section:results}
This section presents the main results of our SED fitting procedure. 
To compare the best-fit parameters for the Milky Way (MW), the Large Magellanic Cloud (LMC), and the Small Magellanic Cloud (SMC), we select, for each galaxy, a subsample composed of those stars that satisfy the following conditions:
\begin{equation}
X_{\mathrm{C\text{-}rich}} < W < 1.7 \ , \ X_{\mathrm{C\text{-}rich}} = 0.7 + 0.15  (M_K + 7.65)^2 \ ,
\end{equation}
where $W$ is the Wesenheit index, defined in this work as the difference between the optical and near-infrared indices, $W = W_{RP, BP-RP} - W_{K_s, J-K_s}$. The individual components are defined as $W_{RP, BP-RP} = G_{RP} - 1.3(G_{BP}-G_{RP})$ and $W_{K_s, J-K_s} = K_s - 0.686(J-K_s)$.
The quantity $X_{\mathrm{C\text{-}rich}}$ defines the carbon star region in the \Gaia-2MASS diagram \citep{lebzelter2018new}.
We further adopt a threshold of $\rm W>1.7$~mag to identify the so-called extreme C-rich stars \citep{lebzelter2018new, nanni2019optical}. This classification, based on optical and near-infrared Wesenheit indices, is distinct from the mid-infrared selection of `Extreme AGB' stars proposed by \citet{blum2006}.
This selection ensures that we compare stars with equivalent photometric properties across the MW, LMC, and SMC, allowing a consistent assessment of their best-fit parameters and any intrinsic differences among the three environments.
In the following section, we refer to this subset as the C-region stars.
Figure~\ref{fig:gaia-2mass-total} shows the location of the C-region stars on the \Gaia-2MASS diagram (carbon stars region).
Table~\ref{tab:median_values_parameters} lists the median values of the best-fit parameters for only the selected C-region stars in the three galaxies, followed by the relative dispersion.
Table~\ref{tab:results_table} shows a preview of the final catalogue, which provides the derived parameters for the entire sample of all stars analysed in this study.

\begin{table}[!ht]
\centering
\caption{Median values of derived parameters for MW, LMC, and SMC C-region stars.}
\label{tab:median_values_parameters}
\resizebox{\columnwidth}{!}{
    \begin{tabular}{lcccccc}
    \hline
    \hline
    Parameter & \multicolumn{2}{c}{MW} & \multicolumn{2}{c}{LMC} & \multicolumn{2}{c}{SMC} \\
    \cline{2-3} \cline{4-5} \cline{6-7} 
     & Median & $\sigma$ & Median & $\sigma$ & Median & $\sigma$ \\
    \hline
    $Mass$ [$\rm M_{\odot}$] & 1.00 & 0.00 & 3.00 & 0.00 & 2.00 & 1.48\\
    $\log{\dot{M}}$ [$\rm M_{\odot}\,\text{yr}^{-1}$] & -7.59 & 0.32 & -7.93 & 0.90 & -7.77 & 1.07\\
    $ \tau_{\rm V}$ & 0.40 & 0.30 & 0.10 & 0.15 & 0.10 & 0.15 \\
    $T_{\rm{dust}}$ [K] & 1000 & 0 & 1000 & 0 & 1000 & 0\\
    $T_{\rm{eff}}$ [K] & 3150 & 222 & 3000 & 297 & 3200 & 445 \\
    C/O & 1.07 & 0.00 & 3.91 & 0.00 & 3.18 & 1.09 \\
    \hline
    \end{tabular}
} 
\tablefoot{
    \tablefoottext{a}{The sigma values quoted represent the intrinsic dispersion of the population (standard deviation), estimated as the median absolute deviation multiplied by 1.4826; note that this dispersion may be asymmetric for some parameters.}
}
\end{table}

\subsection{Effective Temperature}
\label{subsect:Teff}
\begin{figure}[!h]
    \centering
    \includegraphics[scale=0.5]{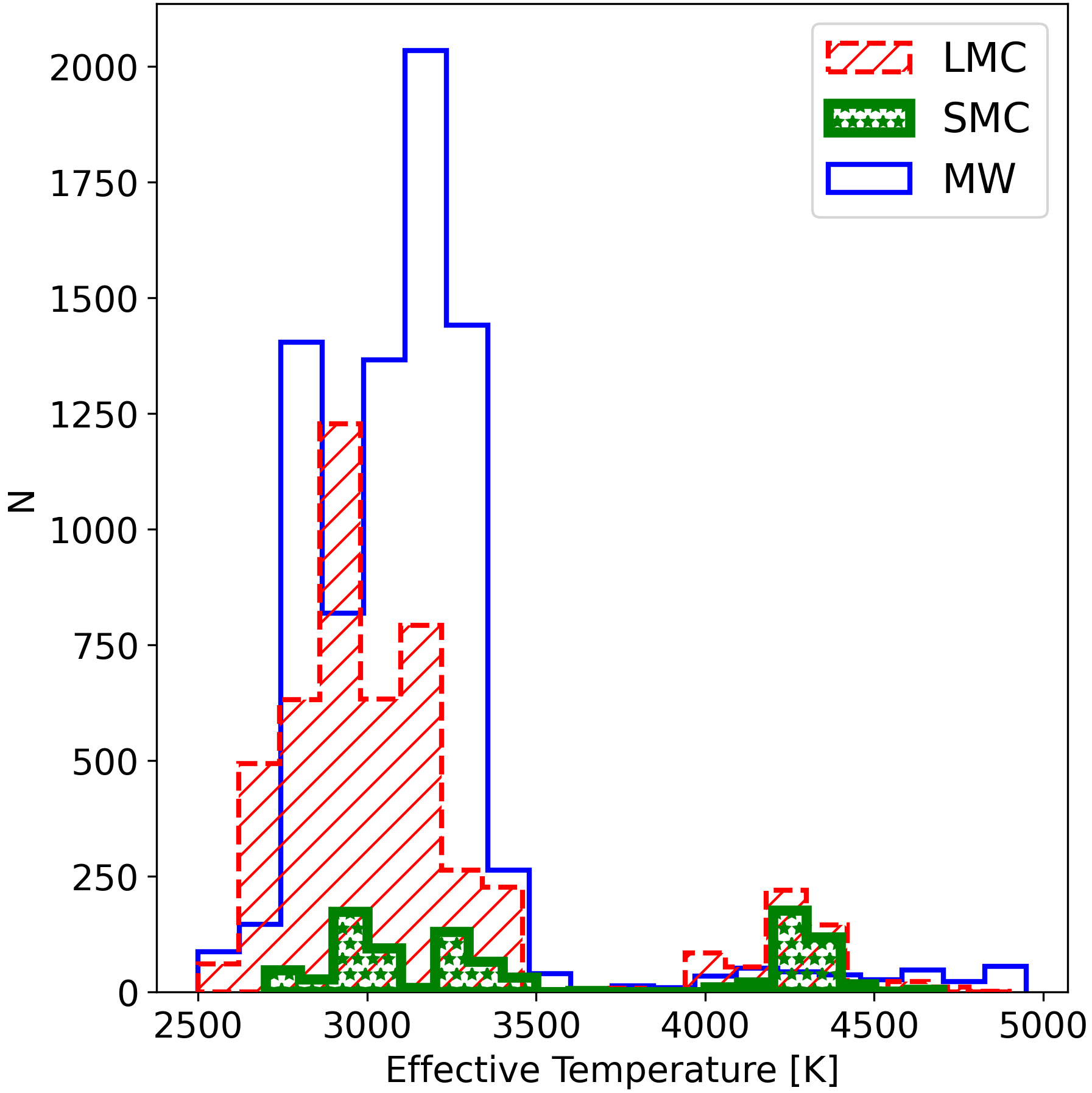}
    \caption{Distribution of $T_\mathrm{eff}$ for C-region stars in the MW, LMC and SMC. Colours and shapes of the three subsamples are showed in the legend.}
    \label{fig:teff_hist}
\end{figure}

The median effective temperature of the entire sample is $T_{\rm eff}=3100$K, with a dispersion of $\sigma T_{eff}=300 \ K$, consistent with the dominant population of carbon stars, as found in many literature studies (e.g.,~\citenop{abia2020carbongaia}). 
Focusing exclusively on the C-region stars subsample within the three populations, Table~\ref{tab:median_values_parameters} summarizes the corresponding statistical properties, while Figure~\ref{fig:teff_hist} shows the distributions for these stars.
Note that the individual measurement uncertainties (also listed in Table \ref{tab:median_values_parameters}) can be primarily affected by the density of the COMARCS grid, which is not equally spaced.
Although the median effective temperatures of the LMC and SMC samples are comparable within the uncertainties, their distributions show an extended high–$T_{\rm eff}$ tail, which is more prominent at lower metallicity. This behaviour is qualitatively consistent with stellar evolution expectations, according to which carbon stars in metal-poor environments can reach slightly higher effective temperatures (e.g.,~\citenop{blanco1983carbonstars}; \citenop{nanni2021dustmetallicity}). 
The comparison between the two Magellanic Clouds and the Milky Way is subject to selection effects, since Gaia samples the carbon star populations differently in these environments, as described in Sect.~\ref{subsec:selection_biases}.

\subsection{Optical depth and mass-loss rate}
\label{subsect:mdot}
\begin{figure}[!h]
    \centering
    \includegraphics[scale=0.5]{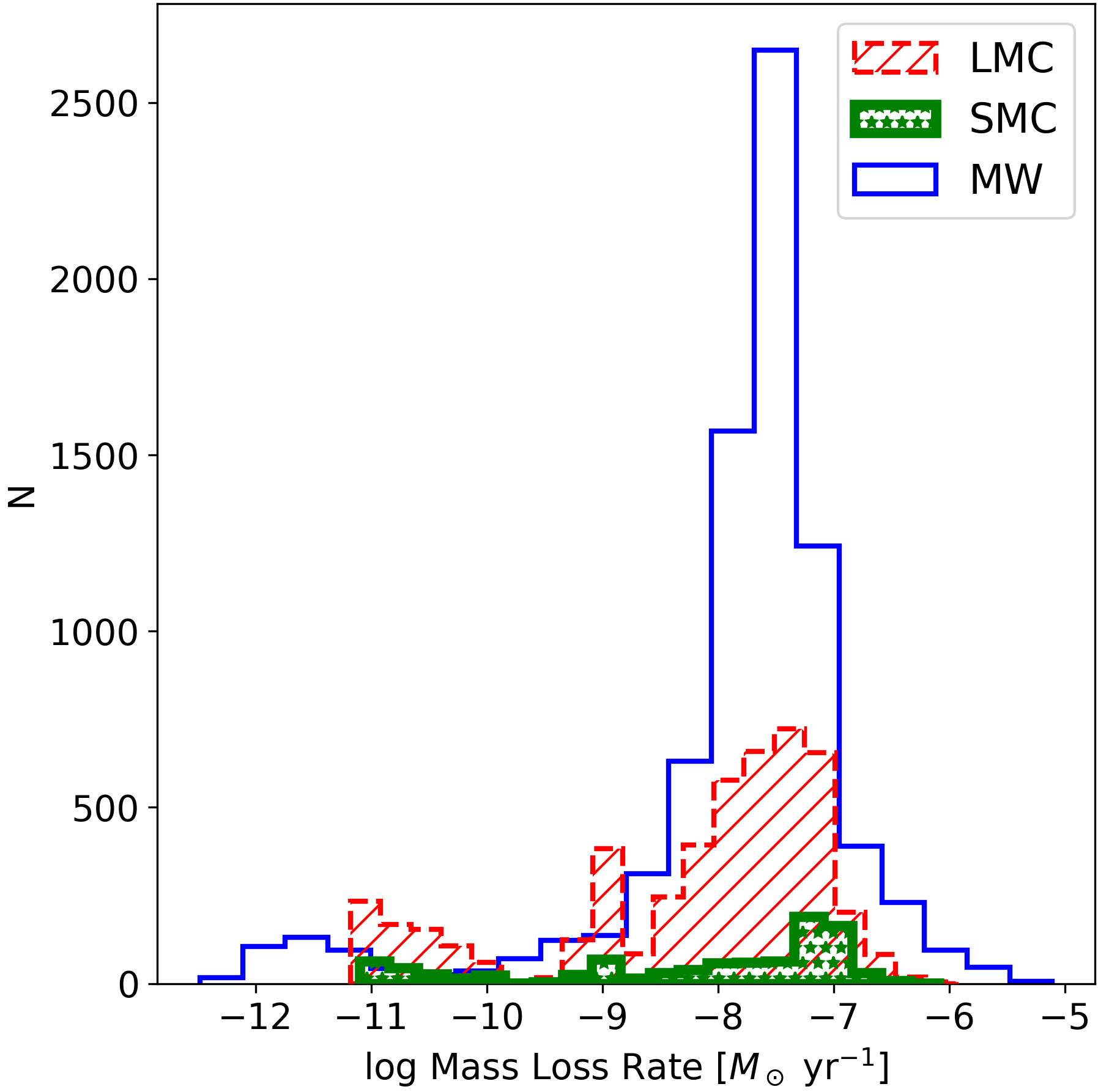}
    \caption{Distribution of mass-loss rates for stars in the MW, LMC and SMC. Colours and shapes of the three subsamples are showed in the legend.}
    \label{fig:massslossrate_hist}
\end{figure}

Figure~\ref{fig:massslossrate_hist} presents the distribution of the mass-loss rates.
These values are fully compatible between both MW and the MCs.
The statistical properties of the mass-loss rates and optical depths for the C-region stars are reported in Table \ref{tab:median_values_parameters}. As shown, both parameters are largely comparable across the three populations, with median mass-loss rates ranging from $\log \dot{M}\approx -7.8$ to -8.2~$M_{\odot}yr^{-1}$ and optical depths $\tau_V$ between 0.1 and 0.4.
Although the MW sources appear to exhibit higher optical depths than their MCs counterparts, these differences are consistent within the estimated uncertainties. Therefore, these trends should be interpreted with caution, given that the distributions also remain sensitive to Gaia selection biases.
We plan to address this issue in a follow-up paper, which will perform a more refined comparison accounting for differences in luminosity, evolutionary stage, and stellar mass.

\subsection{Mass} 
The stellar mass has only a minor influence on the overall shape of the SED, and our fitting procedure is therefore relatively insensitive to this parameter. 
Nevertheless, the derived median values reported in Table \ref{tab:median_values_parameters} are consistent with the expectations for carbon stars (1.5-4~$\rm M_{\odot}$, \citenop{Hofner2018Review}).
Given the discrete nature of the grid, these values should be regarded as indicative rather than quantitatively significant. 
However, the slightly higher average masses inferred for the LMC and SMC are consistent with expectations from stellar evolution at lower metallicities, as determined by \cite{pastorelli2020resolved}.

\subsection{Carbon-to-oxygen ratio}
The carbon-to-oxygen ratio (C/O) is another crucial parameter of the models; the values adopted here are those provided by the COMARCS model grid.
On average, we observe higher C/O ratios in the Magellanic Clouds than in the Milky Way, with the median values presented in Table \ref{tab:median_values_parameters}.
Such a behaviour is qualitatively consistent with expectations for lower-metallicity environments, where the initial oxygen abundance is smaller and fewer dredge-up episodes are required to reach C/O~$>1$.
In addition, stellar evolution models predict more efficient carbon enrichment during thermal pulses at low metallicity \citep{nanni2013, dicriscienzo2013, boyer2013there, abia2020carbongaia}.
However, given the selection effects discussed in Sect.~\ref{subsec:selection_biases}, this result should not be interpreted as direct evidence of a metallicity-driven effect.

Regarding the dust temperature at the inner boundary, for the majority of our entire sample, which is characterized by optically thin envelopes ($\tau_V\lesssim 0.1$), we determine values around $\rm \approx 1000~K$. We note, however, that for such low optical depths, the concept of a single dust temperature is less physically meaningful as the SED is not dominated by the dust emission from a single layer.  Consequently, these values should be treated primarily as boundary conditions for the fit rather than intrinsic properties of the sources.

\begin{table*}[htbp]
\small
    \centering
    \renewcommand{\arraystretch}{1.3} 
    \caption{Properties and derived parameters of the selected stars.}
    \label{tab:results_table}
    \begin{tabular}{l c c c c c c c c c c}
        \hline\hline
        GaiaDR3\_ID & RA & Dec & \ldots & $T_{\mathrm{eff}}$ &$\sigma T_{\mathrm{eff}}$ & $\log \tau_V$ & $\sigma \log \tau_V$ &  $\log \dot{M}$ &  $\sigma \log \dot{M}$ & C/O\\
                    &   (deg) & (deg) & \ldots & (K) & (K) & (dex)  & (dex) & ($M_\odot$\,yr$^{-1}$) & ($M_\odot$\,yr$^{-1}$) & -  \\
        \hline
        5874203376129556480 & 220.542030 & $-63.251443$ & \ldots &  3160 & 38 & -0.22 & 0.14 & $-7.13$ & 0.16 & 1.07143  \\
        5874204235123060352 & 220.539375 & $-63.231429$ & \ldots &  3050 & 34 & -0.31  & 0.32 & $-7.27$ & 0.35 & 1.07143 \\
        \ldots & \ldots & \ldots & \ldots & \ldots & \ldots & \ldots & \ldots & \ldots & \ldots & \ldots \\
        \hline
    \end{tabular}
    \vspace{1mm}
    \tablefoot{
        \tablefoottext{a}{This table is available in its entirety at the CDS. A portion of it is shown here for guidance regarding its form and content.}
    }
\end{table*}

\section{Discussion}
\label{section:discussion}
This work has characterized the stellar and dust properties of the \Gaia\ golden sample of carbon stars. 
As detailed in Sect.~\ref{subsec:selection_biases}, the sample is biased against both dust-obscured and intrinsically faint sources. Since these selection effects impact the MW and MCs differentially, caution is required when performing cross-galaxy comparisons.
In the following, we discuss our findings in the context of analogous studies from the literature.

In this work, we have also explored an alternative approach to the SED-fitting procedure, which adopts a systematic normalization of all the photometry bands to the 2MASS $K_s$ filter. 
This method can ensure an SED-fit without the dependence by the star's distance. 
Appendix~\ref{appendix:methodology_tests} discusses the results of this approach, as well as their comparison to the standard SED-fitting method.

\subsection{Effective temperature}
\label{subsection:teffdiscussion}
As Section~\ref{subsect:Teff} discusses, we derived and compared the effective temperature distributions of MW and MCs carbon stars. 
A comparison of our results with the effective temperatures derived by \citet{andrae2023gaia} reveals a significant discrepancy: their median temperature is $\approx$ 5000~K, significantly higher than ours, with some stars exceeding 10,000~K. 
However, \citet{andrae2023gaia} noted that the GSP-Phot module can provide unreliable parameters for AGB stars. Therefore, our temperature estimates, derived specifically for these cool evolved stars, should be considered more reliable.

\begin{figure*}[!h]
    \centering
    \includegraphics[scale=0.55]{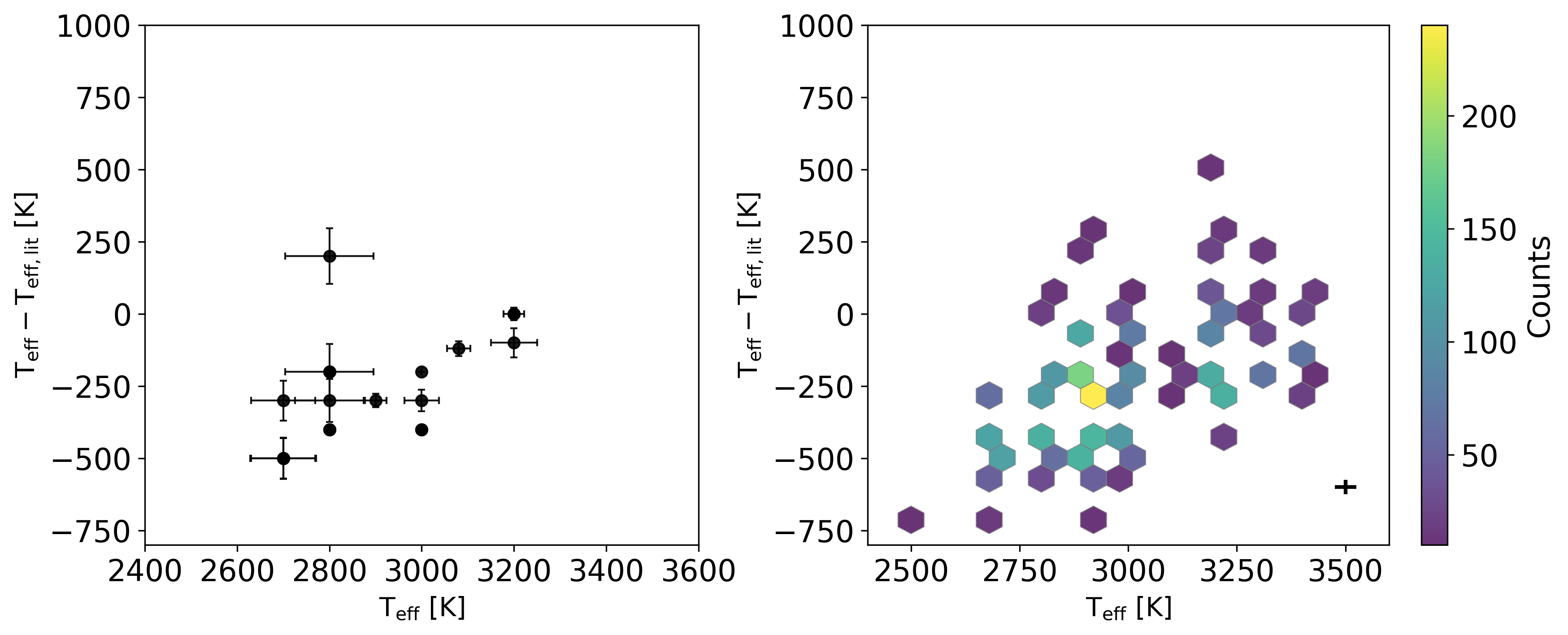}
    \caption{Comparison of the effective temperature values for the common sources between our work and the ones by \cite{groenewegen2009luminosities}, \cite{groenewegen2018luminosities} (left diagram) and \cite{riebel2012mass} (right diagram). The data are shown using hexagonal density binning with a grid size of 40 and a threshold count of 10.}
    \label{fig:teff-literature}
\end{figure*}

We further compared our results with previous SED-fitting studies of carbon stars, namely \citet{groenewegen2009luminosities}, \citet{groenewegen2018luminosities}, \citet{groenewegen2022wise}, and \citet{riebel2012mass}.
Figure~\ref{fig:teff-literature} directly compares our results and with those of the literature.
These works were selected because they share stars with our sample, adopt similar physical assumptions, and cover both the MW and the MCs, while differing in their treatment of circumstellar dust and radiative-transfer methods. 
In total, we identified 3 stars in common with \cite{groenewegen2009luminosities}, 21 with \cite{groenewegen2018luminosities} and 3,839 with \cite{riebel2012mass}, enabling a robust comparison.

Overall, we find good agreement with these SED-fitting studies of carbon stars. For all stars in common, the differences in effective temperature are within $\pm500$~K (see the left panel of Fig.~\ref{fig:teff-literature}), and a mild correlation is observed between the two $T_{\rm eff}$ estimates, with the coolest stars in our sample tending toward higher temperatures in the literature values.

All the aforementioned SED-fitting studies of carbon stars adopt COMARCS atmospheric models for the stellar component, but differ in their treatment of circumstellar dust and in the radiative-transfer codes employed: \citet{riebel2012mass} used 2DUST, whereas the others relied on DUSTY or DUSTY-based codes (e.g., MoD; \citealt{groenewegen2012mod}). Differences in $T_{\rm eff}$ may therefore arise from several factors, including the choice of radiative-transfer code, grid resolution, and photometric datasets.

Additional sources of uncertainty include extinction and distance. Inaccurate distances can bias the best-fit models, as the fitting procedure compensates to reproduce the observed photometry and can also affect the inferred extinction. This effect is particularly relevant for MW stars, whereas distances and extinction values for MCs stars are well constrained and largely consistent across comparable studies. Finally, differences in the adopted dust optical constants can significantly modify the emerging SED shape, further affecting the derived effective temperatures.

\subsection{Mass-loss rate}
The mass-loss rate distributions of the Milky Way and the Magellanic Clouds show consistent results (see Sect.~\ref{section:results} and Fig.~\ref{fig:massslossrate_hist}).
For carbon stars, the dependence of mass-loss on metallicity is not straightforward. 
Since carbon is produced internally through dredge-up episodes, theoretical models and observations suggest that carbon stars can drive strong winds even at low metallicity \citep[e.g.,~][]{vanloon2000masslossrelatio,groenewegen2007luminosities, Hofner2018Review}. 
The intrinsic mass-loss rate of C-stars is primarily driven by the subsequent formation of amorphous carbon (amC) dust, the key mechanism responsible for wind acceleration \citep{nanni2021dustmetallicity}.
From a theoretical perspective, metal-poor environments are expected to favour higher surface C/O ratios, owing to the lower initial oxygen abundance and to the increased efficiency of the third dredge-up at low metallicity.
This condition can, in principle, enhance the circumstellar opacity and facilitate mass loss.
However, such expectations apply to stars at comparable evolutionary stages and luminosities, and do not necessarily translate into observable differences when heterogeneous samples are compared.

Two main differences emerge between the MW and MC samples: considering the whole catalogue, the MW contains a larger number of stars with lower mass-loss rate and a slightly higher fraction of stars with higher mass-loss rate compared to the MCs (see Fig.~\ref{fig:massslossrate_hist}).
These differences are almost certainly driven by selection effects rather than intrinsic physical variations, as described in Sect.~\ref{subsec:selection_biases}. 

Moreover, as mentioned above, a robust assessment of metallicity-related effects on mass loss would require a comparison between stars matched in fundamental parameters, such as luminosity and evolutionary stage.
Such an analysis, while enabled by the large sample and the availability of \Gaia\ parallaxes, is beyond the scope of the present work.

\begin{figure*}
    \centering
    \includegraphics[scale=0.55]{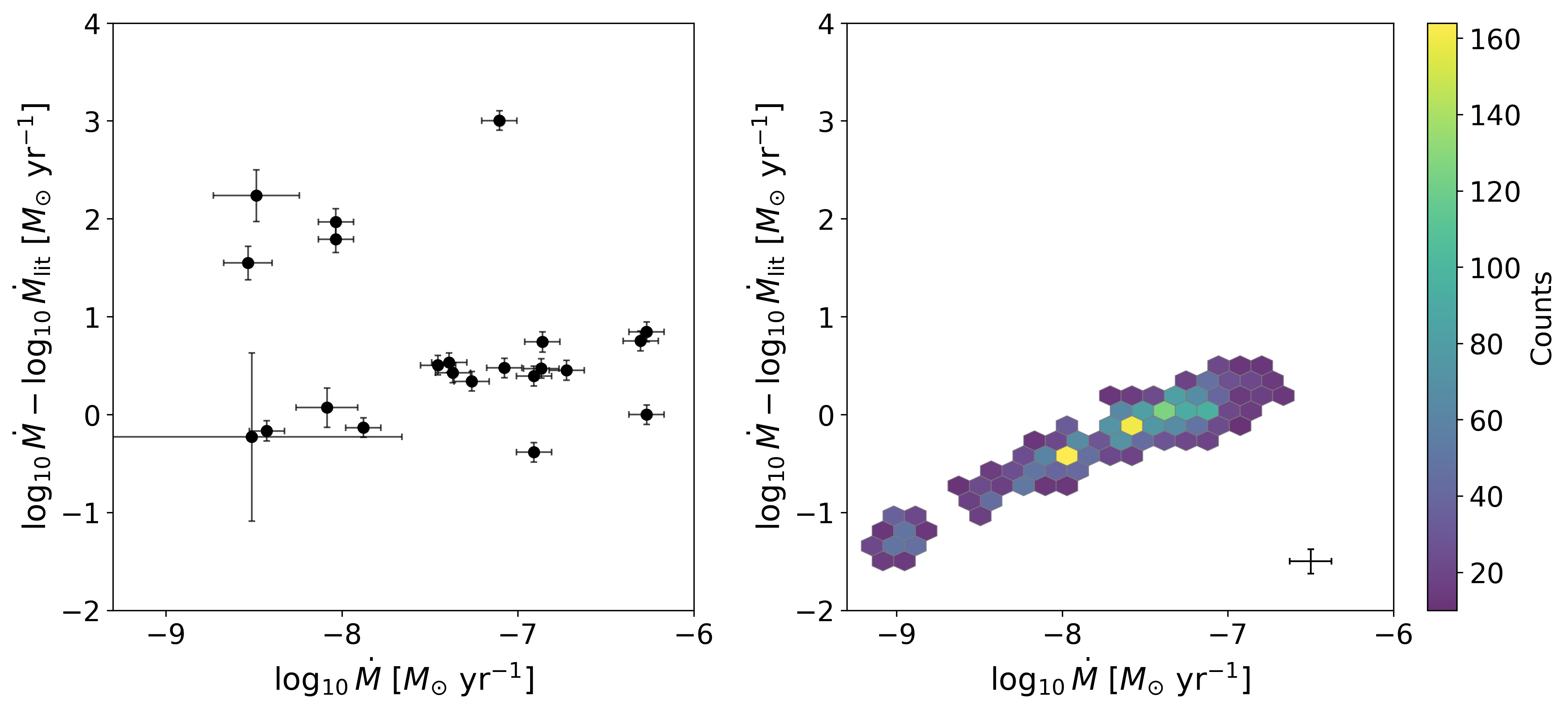}
    \caption{Comparison of the gas mass-loss rate values for the common sources between our work and the ones by \cite{groenewegen2009luminosities}, \cite{groenewegen2018luminosities} (left diagram) and \cite{riebel2012mass} (right diagram). Uncertainties plotted are only the ones derived in this study. The cross on the bottom of the right panel represents the average uncertainty. We note that the average uncertainty depends on the mass-loss regime, as it is clear on the left panel.}
    \label{fig:mdot_comparison}
\end{figure*}

We also compared our estimated mass-loss rate with the same set of studies used for the $T_\mathrm{eff}$ comparison (see Sect.\ref{subsection:teffdiscussion}).
The results, shown in Fig.~\ref{fig:mdot_comparison}, do not reveal any clear systematic offset across the full range of mass-loss regimes.

In particular, when compared with the values from \citet{riebel2012mass}, we find an excellent agreement, with almost all stars lying within 1 dex of the zero-offset line.
A comparison with \citet{groenewegen2009luminosities,groenewegen2018luminosities} also shows a generally good consistency, with an average offset of about 0.5 dex and no evident trends.
We note that the apparent trend with mass-loss rate visible in Fig.~\ref{fig:mdot_comparison} likely reflects the increasing sensitivity of SED-based $\dot{M}$ determinations to modelling assumptions at both low and high optical depths, rather than a genuine physical discrepancy.
While a 1 dex difference represents an order of magnitude, such a spread is not unexpected in the context of AGB mass-loss modelling, given the high sensitivity of the results to the adopted model parameters.
For instance, five stars in common exhibit, in our study, mass-loss rates that are 2–3 dex higher than the values reported in those studies (see Fig.\ref{fig:mdot_comparison}).
For these five objects, we find differences in bolometric luminosity of approximately 30–50\% relative to our values. 
It is worth noting that SED fitting becomes intrinsically uncertain at low optical depth.
In this regime, even modest variations in the adopted luminosity (or in any parameter that affects the SED normalization) can shift the inferred optical depth by a significant amount, which in turn propagates into large changes—up to several dex—in the derived mass-loss rate.
This sensitivity is an important limitation in modelling low-$\tau$ carbon stars, and likely contributes to the offsets observed for these five objects.
Furthermore, as demonstrated by \citet{groenewegen2018luminosities}, the choice of different optical constants for amorphous carbon grains (e.g.,~\citealt{rouleau1991amc} vs. \citealt{Zubko1996}) can independently shift the derived $\dot{M}$ up to a factor of several.

\citet{nanni2019mass} provide a more systematic assessment of the possible caveats, highlighting how several modelling assumptions can easily bias the final estimates.
Beyond the choice of the dust optical constants, other ingredients of the models may significantly affect the outcome, such as the adopted initial expansion velocity or the neglect of dust–gas drift, which can alter the relation between gas and dust densities.
Another fundamental ingredient that can have a strong impact on the results, in particular on the derived mass-loss rates, is the treatment of the wind dynamics. 
Even the assumed coefficients for amorphous carbon, as mentioned before, or the prescription for the seed nuclei abundance, can play a role. 
Additional uncertainties come from possible deviations from spherical symmetry of the model, as well as from photometric variability, which introduces extra scatter in the SED fitting and is taken into account in this work (see Sect.~\ref{section:lightcurves} and Appendix~\ref{appendix:lightcurve}). 
Taken together, these factors imply that mass-loss rates and related quantities should always be considered within the context of the adopted modelling framework, rather than as absolute values.
Possible discrepancies might also be attributed to differences in the SED fitting parameters or wavelength coverage. 

In addition to the star-by-star comparison with literature values, we examined the global behaviour of our results in the luminosity–mass-loss rate ($L$–$\dot{M}$) diagram, shown in Fig.~\ref{fig:mdot-lum-literature}.
For the LMC, SMC, and MW, we find that our stars (black points) occupy the same range of luminosities and mass-loss rates as reported by previous studies, confirming the overall consistency of our determinations.
A slight selection bias from the \Gaia\ sample is evident, however, as both the MCs and MW samples do not reach the most extreme mass-loss regimes, as the stars studied by \cite{groenewegen2009luminosities}, \cite{groenewegen2018luminosities} and \cite{groenewegen2022wise}.
Moreover, our mass-loss estimates are consistent with expectations from theoretical stellar evolution models (see \citenop{pastorelli2019smc} and \citenop{pastorelli2020resolved}). 

\subsection{\Gaia-2MASS diagram}
\begin{figure*}[!h]
    \centering
    \includegraphics[scale=0.45]{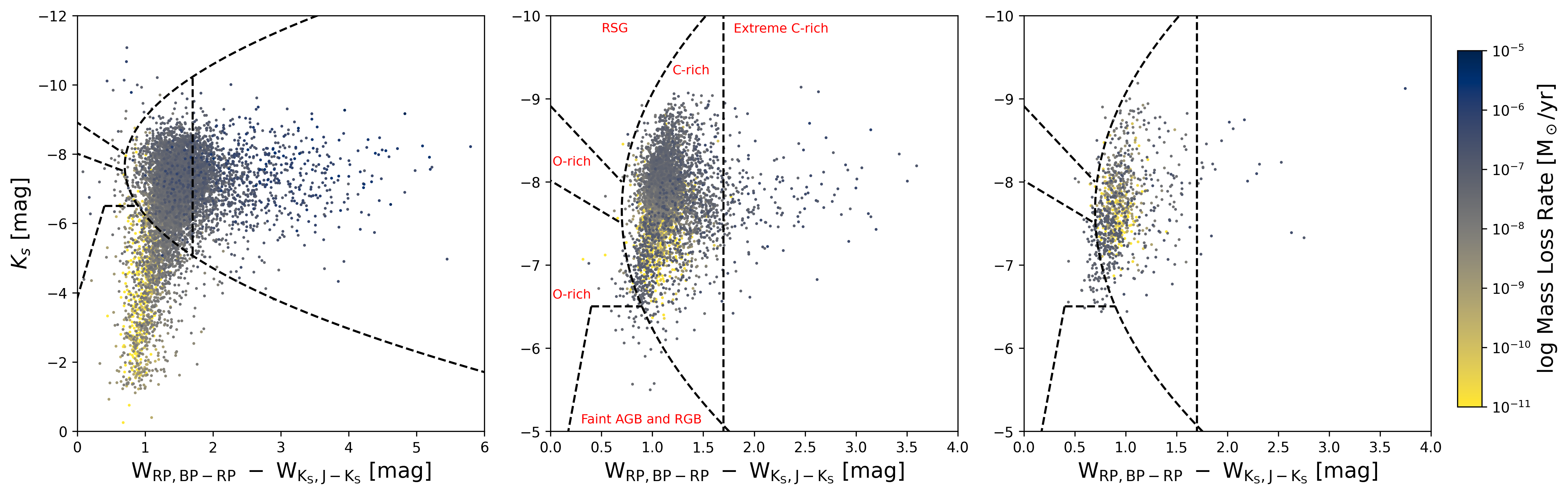}
    \caption{From left to right: MW, LMC and SMC stars. \Gaia-2MASS diagram coloured by the optical depths $\tau_{V}$ values derived in our study. The dotted lines represent the observational-derived lines delimiting the different regions occupied by different types of AGB stars, and are defined as by \cite{lebzelter2018new}. In the left panel, each region has been labelled accordingly with red text. The stars inside the C-rich region are the ones defined in this paper are C-region stars.}
    \label{fig:gaia-2mass-total}
\end{figure*}

The \Gaia–2MASS diagram (see Sect.~\ref{section:results} for its definition) provides further insight into the catalogue adopted and our findings.
Introduced by \citet{lebzelter2018new}, it combines optical and near-infrared photometry to provide an efficient diagnostic for the chemical type and evolutionary status of AGB stars. Figure~\ref{fig:gaia-2mass-total} shows this diagram for our entire sample, colour-coded by the mass-loss rate derived from SED fitting.
Magellanic Clouds carbon stars populate the expected region, with the dustiest and most evolved objects—characterized by higher $\dot{M}$ values—progressively shifting toward redder colours, and in general all occupying the same region in the diagram. 

\begin{figure*}
    \centering
    \includegraphics[scale=0.35]{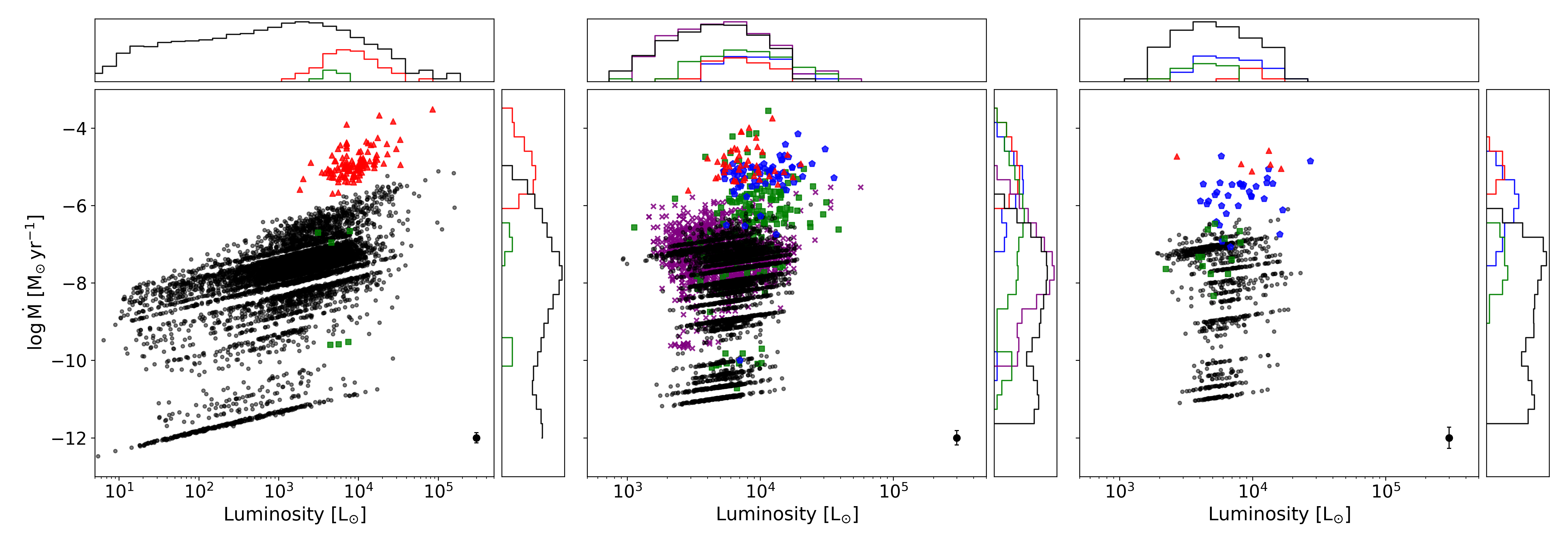}
    \caption{From left to right, MW, LMC and SMC stars. Luminosity against mass-loss rate values for stars in our sample (black dots), \cite{groenewegen2022wise} (red triangles), \cite{groenewegen2009luminosities} (blue pentagons), \cite{groenewegen2018luminosities} (green squares) and \cite{riebel2012mass} (purple crosses). The black point on the bottom right of each panel represents the average uncertainty of our estimated mass-loss rates for each galaxy.}
    \label{fig:mdot-lum-literature}
\end{figure*}

The MW sample behaves similarly. The majority of MW stars are located in the same carbon-rich region as their MC counterparts, confirming their C-rich classification. 
However, the MW population shows a broader spread, including a tail toward redder Wesenheit values with higher $\dot{M}$, suggesting the presence of more extreme-carbon stars, which are characterised by an $\dot{M}>10^{-6}~\mathrm{M_{\odot}yr^{-1}}$ \citep{nanni2019mass}.
Additionally, several MW sources occupy a low-luminosity region where we expect to see a mix of faint AGB and RGB stars. 
These sources likely correspond to less-evolved, dust-poor carbon stars, whose photometric properties have yet to be significantly affected by mass loss. 
This overall distribution confirms that the Gaia–2MASS diagram not only distinguishes between O-rich and C-rich AGB populations but also traces evolutionary effects such as dust formation and mass-loss enhancement.
In addition, we highlight that the large size of our sample allows us to statistically trace the general distribution of the dust and stellar parameters of these carbon stars. The grid-like appearance of our results in Fig.~\ref{fig:mdot-lum-literature} reflects the discrete sampling of optical depth values adopted in our model grid. This does not affect the global trends, but it explains why our points appear more regularly spaced than those from other studies.
A rigorous decontamination would require a dual-grid fitting approach (C-rich vs O-rich), which is beyond the scope of this work.
\subsection{Variability of the Carbon stars}
\label{subsect:variability}
\begin{figure}[!h]
    \centering
    \includegraphics[width=0.9\linewidth]{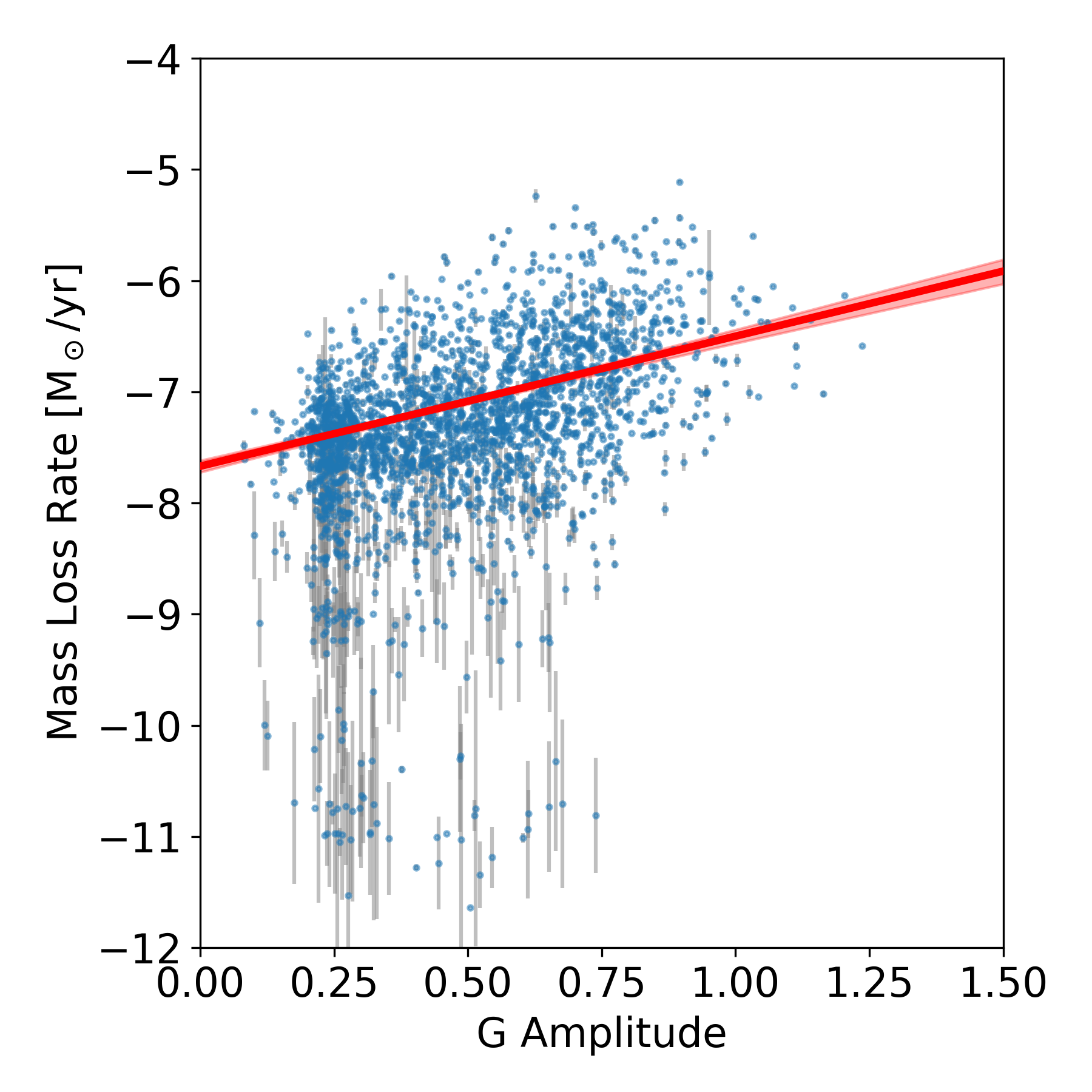}
    \caption{Scatter plot of \Gaia\ G mean magnitude derived from our light curve fit method (see Sect.~\ref{section:lightcurves}) and mass-loss rates from our SED fit. The red dotted line is a linear fit between amplitude and mass-loss.}
    \label{fig:mdot-amplitudeG}
\end{figure}

Since the pulsations of carbon stars are the main driver of their mass-loss processes, we expect a connection between their photometric variability and the circumstellar dust content.
Indeed, we find a clear correlation between the mass-loss rate derived from our models and the variability amplitude measured in the \Gaia\ $G$ band, in the sense that stars with larger amplitudes tend to show higher $\dot{M}$.
The fitted relation we derived is $G_{\rm Amplitude} = s \cdot \log \dot{M} + h$, with $s = 1.17 \pm 0.1$ and $h = -7.67 \pm 0.1$.
In the linear fit, measurements with lower uncertainties were assigned higher weights.
At the same time, we note that the dispersion in mass-loss rate at a given pulsation amplitude is substantial, reaching approximately two orders of magnitude, higher than the intrinsic uncertainty.
This large spread indicates that pulsation amplitude alone is not sufficient to uniquely determine the mass-loss rate.
Other stellar properties, such as luminosity, evolutionary stage (the progression along the TP-AGB), and chemical composition, are likely to contribute significantly to the observed scatter.
Within these limitations, the observed trend likely reflects the fact that stronger pulsations can facilitate the injection of material into the stellar envelope, thereby favouring dust formation and enhanced mass loss.
A similar qualitative behaviour has been reported for oxygen-rich Mira variables, where optical variability amplitudes correlate with the intensity of SiO maser emission, interpreted as a consequence of pulsation-driven radiative pumping \citep{alcolea1990sio}.
Figure~\ref{fig:mdot-amplitudeG} illustrates both the overall trend and the large intrinsic scatter in the relation between mass-loss rate and $G$-band amplitude.

\subsection{Expectation from theoretical models}
\cite{aringer2019carbon} showed that spectra and colours of C stars are most sensitive to C/O ratio rather than to the overall metallicity, and they are also strongly affected by $T_\mathrm{eff}$ and molecular line saturation. 
Spectral markers of nitrogen and oxygen enrichment (e.g. CN, $\rm C_2$, and CO features) also depend on $T_\mathrm{eff}$, becoming more evident at higher $T_\mathrm{eff}$ values. 
In particular, at higher $T_{\mathrm{eff}}$, variations in O and N abundances have a stronger and more observable impact on the spectra, as molecular absorption becomes less saturated.
These trends broadly agree with expectations that C star spectra are less sensitive to metallicity and $T_\mathrm{eff}$ than M star spectra, which are dominated by metallicity-linked molecular species like TiO.

\section{Conclusions}
\label{section:conclusion}
In this work, we have presented a homogeneous SED-fitting analysis of a large sample of carbon stars in the Milky Way and the Magellanic Clouds, combining Gaia DR3, 2MASS and WISE photometry with the radiative transfer code DUSTY. We derived effective temperatures, luminosities, optical depths and mass-loss rates for 14,747 sources. 
The results confirm the general picture of carbon-rich TP-AGB stars as efficient mass-losing objects, with typical average effective temperatures at 3150 K and a standard deviation of 430 K. 
Mass-loss rate values span the range $10^{-11}$–$10^{-4}\,~\rm M_\odot\,{\rm yr^{-1}}$. 
Specifically regarding the LMC, we find a distribution of mass-loss rates consistent with the theoretical predictions calibrated for this galaxy by \cite{pastorelli2020resolved}.

We find that stars with larger variability amplitudes generally show higher mass-loss rates, supporting the hypothesis that pulsations play a major role in levitating material into the dust-formation zone. 
However, as noted in the analysis (see Sect.~\ref{subsect:variability}), a large spread in $\dot{M}$ for a given amplitude is present, indicating that other factors also play a significant role in driving the mass loss.
The distributions of effective temperature and mass loss differ across environments. In particular, the lower-metallicity samples exhibit an extended high-$T_{\rm eff}$ tail, rather than a strong systematic shift in the median values, in qualitative agreement with expectations from metallicity-dependent stellar evolution.
However, selection biases may affect this result. The comparison with previous studies reveals broad agreement in the trends but also  differences of up to some orders of magnitude in the absolute values of $\dot{M}$. 
These discrepancies reflect the strong sensitivity of the derived quantities to the adopted physical assumptions.

As emphasized by \citet{nanni2019mass}, the determination of mass-loss rates through SED fitting inevitably depends on the models. 
The choice of optical constants for amorphous carbon grains, the treatment of wind dynamics, or the assumed gas-to-dust ratio and expansion velocity can significantly affect the derived $\dot{M}$. 
Additional uncertainties arise from spherical symmetry and from photometric variability. 
A major role can also be played by the treatment of the dynamic interaction between gas and dust around the AGB star.
Nevertheless, the homogeneous methodology adopted in this work ensures that all sources have been analysed within a consistent modelling framework. This enables reliable relative comparisons among stars in different environments, even if the absolute values depend on the adopted physical prescriptions.

Despite these caveats, our analysis delivers an extensive and homogeneous catalogue of stellar and circumstellar parameters for carbon stars in the Gaia DR3 Golden Sample. These results provide a valuable reference for future studies of carbon-star properties based on Gaia data. It may also serve to refine the criteria for the identification and classification of carbon stars in forthcoming Gaia data releases, including DR4 and DR5.

Future progress will benefit from extending the fitting procedure to larger and more diverse samples, ideally encompassing all available carbon-star catalogues within a consistent and homogeneous framework. 
It will also be essential to use machine-learning methods to create continuous model grids, in order to reduce the dependence of the results on the grid itself, and to adopt methods that account for the time-dependent behaviour of pulsating AGB stars. In this context, fitting SEDs at different phases of the pulsation cycle would enable a more complete characterization of the variability in stellar parameters. A particularly promising direction involves the development of time-dependent hydrodynamic models that include the effects of pulsation, together with the application of Bayesian statistical techniques to quantify uncertainties and correlations among the derived quantities. 
Moreover, improved distances and extinction values will play a crucial role in the determination of the parameters, especially for the MW sample.
These advances will enable a more robust use of AGB carbon-star samples in studies of stellar evolution and dust production in galaxies.

\section*{Data availability}
The results presented in Table \ref{tab:results_table} are only available in electronic form at the CDS via anonymous ftp to \url{cdsarc.u-strasbg.fr} (130.79.128.5) or via \url{http://cdsweb.u-strasbg.fr/cgi-bin/qcat?J/A+A/}.
The ATHENA-C model grid, described in Sec.~\ref{section:models} (or Table \ref{tab:model_grid}), is openly available on Zenodo at \url{https://doi.org/10.5281/zenodo.18418537}.

\begin{acknowledgements}
This project has received funding from the European Union’s Horizon 2020 research and innovation programme under the Marie Skłodowska-Curie grant agreement No 101072454, as part of the Milky Way-\Gaia\ Doctoral Network (\url{https://www.mwgaiadn.eu/}).
The authors, especially A.L., would like to thank S. Srinivasan for the important and useful conversations regarding the statistical treatment of the results and uncertainties; A. Bonanos for her guidance and support during the course of the project; and C. Kalup for the important discussion on the complex mosaic of available Galactic extinction maps.

The contribution by MT and GP is funded by the European Union – NextGenerationEU and by the University of Padua under the 2023 STARS Grants@Unipd programme (``CONVERGENCE: CONstraining the Variability of Evolved Red Giants for ENhancing the Comprehension of Exoplanets'').

GC acknowledges financial support from European Union—Next Generation EU, Mission 4, Component 2, CUP: C93C24004920006, project ‘FIRES'.

This work has made use of data from the European Space Agency (ESA) mission {\it Gaia} (\url{https://www.cosmos.esa.int/gaia}), processed by the {\it Gaia} Data Processing and Analysis Consortium (DPAC, \url{https://www.cosmos.esa.int/web/gaia/dpac/consortium}). Funding for the DPAC has been provided by national institutions, in particular the institutions participating in the {\it Gaia} Multilateral Agreement.

This research has made use of the VizieR catalogue access tool \citep{ochsenbein2000vizier}, CDS, Strasbourg Astronomical Observatory, France (DOI : 10.26093/cds/vizier).

This research has made use of the Spanish Virtual Observatory (https://svo.cab.inta-csic.es) project funded by MCIN/AEI/10.13039/501100011033/ through grant PID2020-112949GB-I00 \citep{rodrigo2020svo}.

This research has used data, tools and materials developed as part of the EXPLORE project that has received funding from the European Union’s Horizon 2020 research and innovation programme under grant agreement No 101004214.

We wish to thank the "Summer School for Astrostatistics in Crete" for providing training on the statistical methods adopted in this work.

\end{acknowledgements}

\vspace*{-0.5cm} 
\bibliographystyle{aa} 
\bibliography{main_bibliography.bib} 
\begin{appendix}
\section{Lightcurve analysis}
\label{appendix:lightcurve}
To analyse the variability of our sample stars, we developed a dedicated Python code that processes time-series photometry and reconstructs phase-folded light curves across multiple filters. The code follows a modular design, ensuring that each step of the analysis is fully reproducible and easily adaptable to different datasets. The input consists of \Gaia\ DR3 epoch photometry and, when available, complementary data from OGLE. These measurements are assembled into a dedicated \texttt{TimeSeries} object that standardizes the data structure by storing observation times, magnitudes with their associated uncertainties, and the corresponding photometric bands. This approach enables the consistent treatment of sources with heterogeneous sampling and multi-band observations within a unified framework.

The first task of the code is the determination of the variability period. For this purpose, we implemented the multiband LombScargle periodogram, which simultaneously fits all available photometric bands and maximizes the likelihood of a common variability period, while allowing for differences in amplitude and phase among the filters. 
The algorithm provides as output the best-fit period, its uncertainty, and the complete periodogram for diagnostic purposes. 
For those stars in common with the OGLE survey, we verified that the periods derived by our code are fully consistent with the OGLE values, thus confirming the reliability of our implementation.

\begin{figure}[!h]
    \centering
    \includegraphics[scale=0.42]{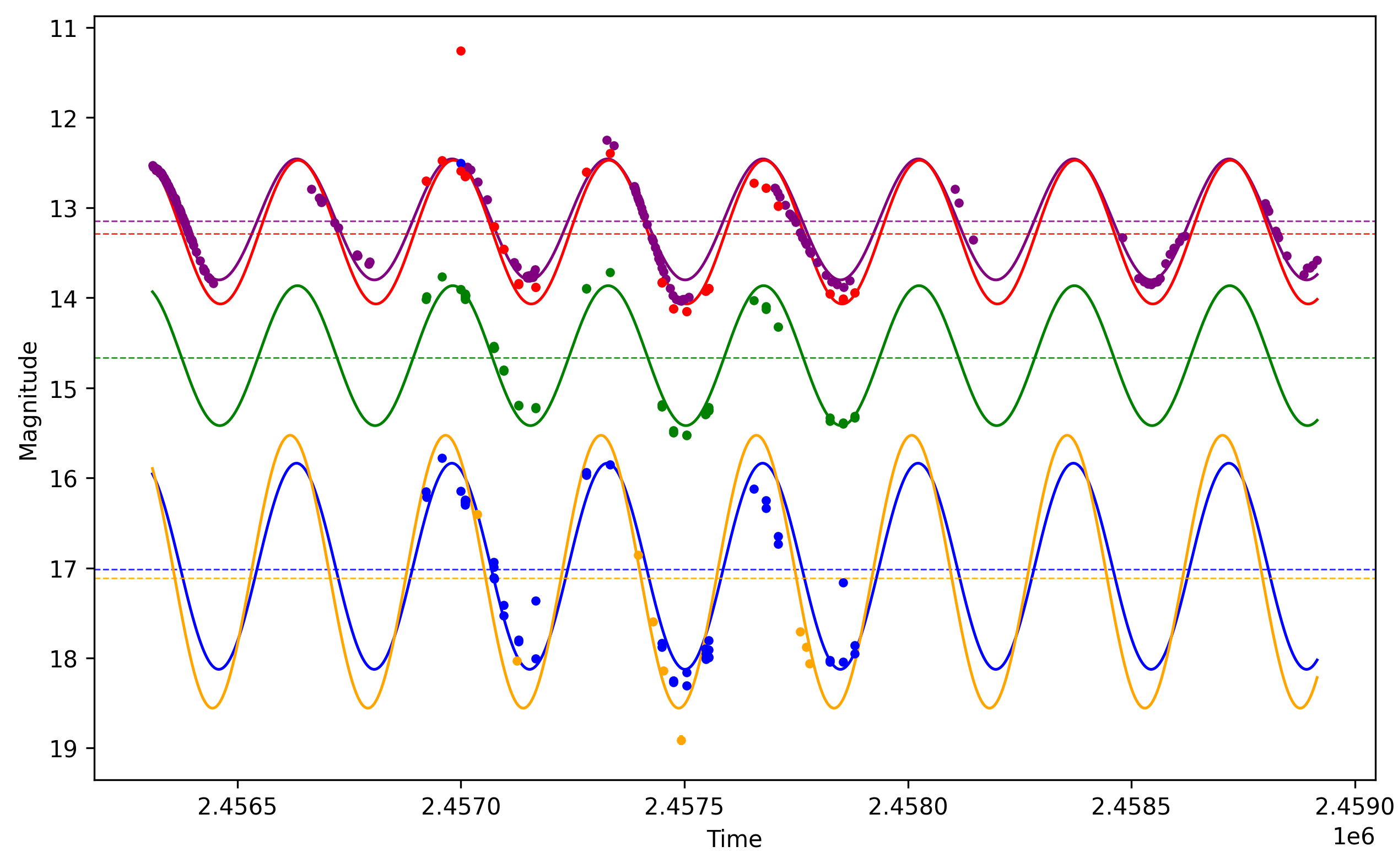}
    \caption{Observed multiband and multi-period photometry for the stars Gaia DR3 2935152926080940544 with the best-fit light curve models. The points represent the observed magnitudes, and the curves represent the best-fit light curve for each photometry band. The horizontal dashed lines are the mean magnitudes for each best-fit model of each band. To create the multiband light curve model for this star, we used OGLE bands I (purple) and V (yellow), and \Gaia\ bands G (green) BP (blue) and RP (red).}
    \label{fig:lightcurve_GaiaDR3_2935152926080940544}
\end{figure}

Once the period is determined, the code reconstructs the light curve model. The data are folded with the derived period and fitted with a multi-harmonic Fourier series across all available bands. This step allows us to produce both a global model of the variability and individual-band reconstructions. 
Figure ~\ref{fig:lightcurve_GaiaDR3_2935152926080940544} shows an example of a light curve multiband reconstruction.
The resulting models are used to investigate the variability properties of the stars and to provide homogeneous light curve products.
We decided to fit with one single period to each star in order to be more accurate for the Miras light curves.
The code also has the option for a multi-periodic.

The outputs include the best-fit periods with uncertainties, the model light curves for each band, phase-folded plots of the data and models, classification flags (e.g., Miras, other LPVs), and diagnostic statistics such as $\chi^{2}$ and residual scatter. These products are collected in structured tables and figures, which form the basis of the final catalogue used in this work.

\section{Testing the chi-squared methodology}
\label{appendix:methodology_tests}
During the SED-fitting analysis, we tested several alternative approaches to assess their impact on the derived stellar parameters. Among these, the most relevant test concerned the normalization of both observed and model fluxes to the $K$ band. In this approach, the $K$ band is fixed, and the fitting becomes sensitive primarily to the shape of the SED rather than to its absolute scaling. 
\cite{yang2023evolved} provide an example of an SED fitting study that adopts this methodology.

\begin{figure}
    \centering
    \includegraphics[width=0.8\linewidth]{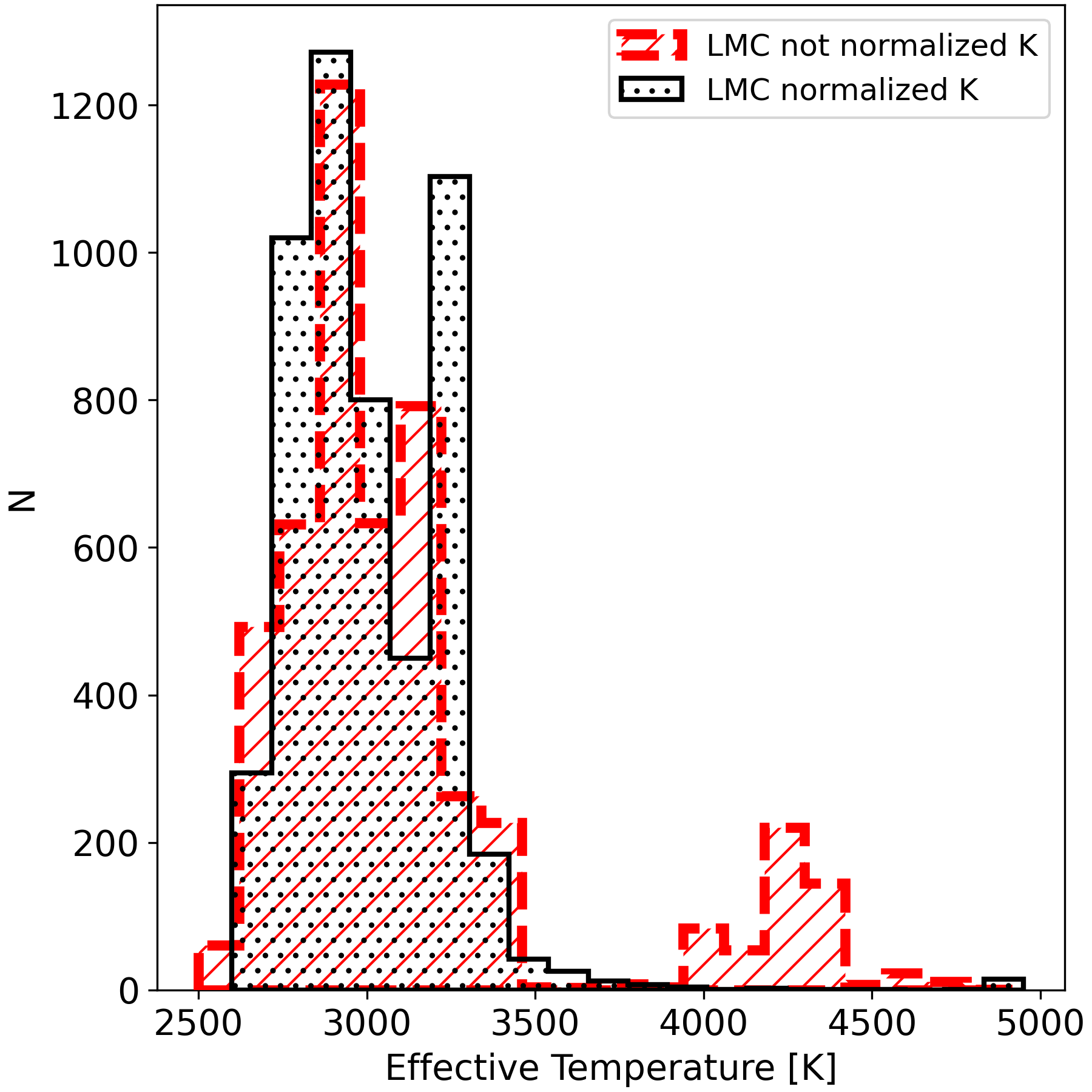}
    \caption{Histogram of distribution of the effective temperature for only LMC stars, derived by the classical SED fitting that we applied in this work (red distribution) and the SED fitting using both observations and models normalized to the K filter (black).}
    \label{fig:difference_normK_notnorm_Teff}
\end{figure}

\begin{figure}
    \centering
    \includegraphics[width=0.8\linewidth]{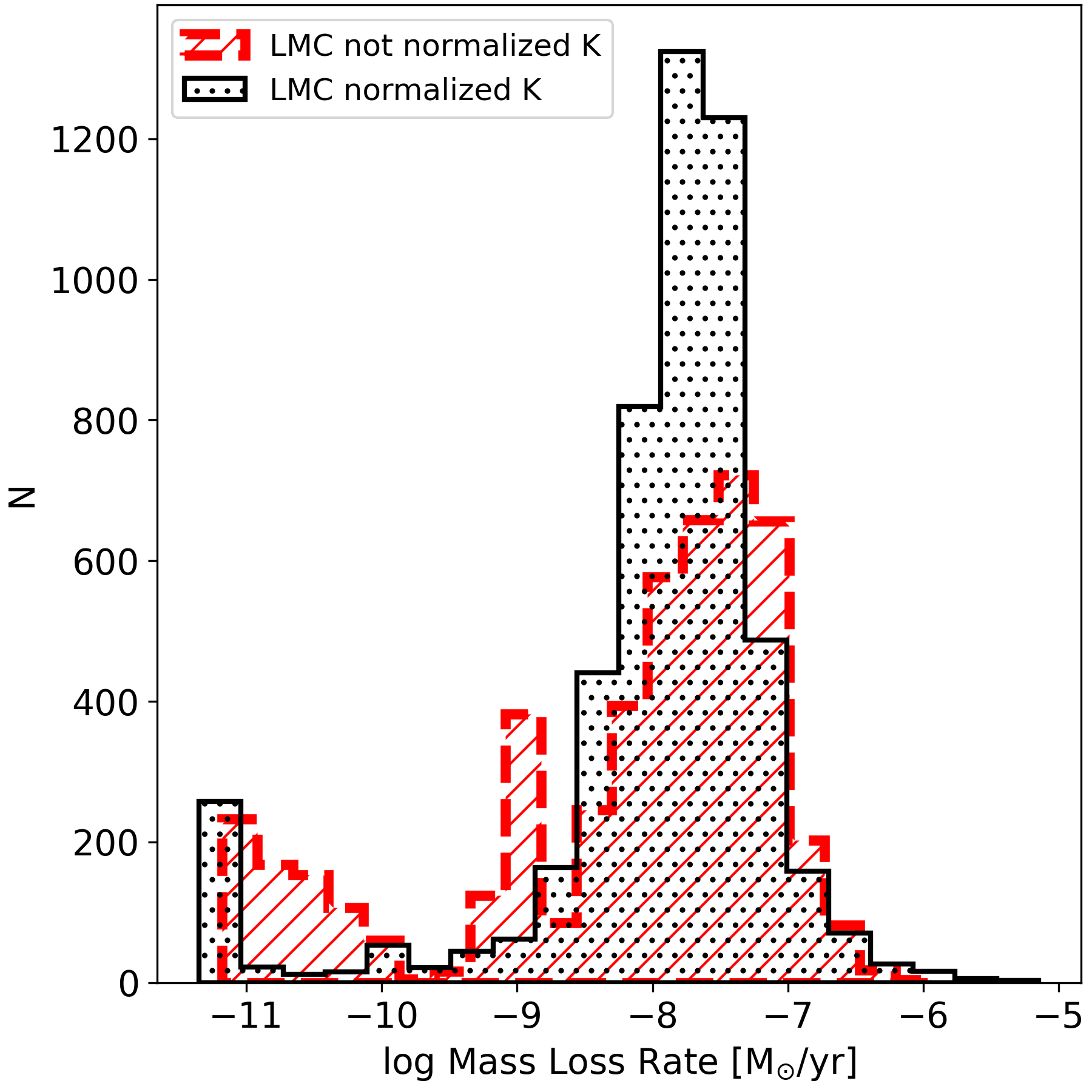}
    \caption{Histogram of distribution of the mass-loss rates for only LMC stars, derived by the classical SED fitting that we applied in this work (red distribution) and the SED fitting using both observations and models normalized to the K filter (black).}
    \label{fig:difference_normK_notnorm_mass-loss}
\end{figure}

The results presented in Fig.~\ref{fig:difference_normK_notnorm_Teff} show that the effective temperature distribution obtained with the $K$-band normalization is compatible, within the uncertainties, to the one derived with classical method.
The only notable difference in the temperature distribution is the presence of hotter stars with the classical method. The same stars studied with the K-normalization method have a temperature compatible with the bulk of the sample.
The normalization tends to yield lower optical depths and, consequently, lower mass-loss rates compared to the classical method, described in Sect.~\ref{section:results}.
In particular, as shown in Fig.~\ref{fig:difference_normK_notnorm_mass-loss}, it is clear that K-normalized results provide generally lower mass-loss values and more sparse towards the low-end of the mass-loss distribution, while the results derived with the general classical method provide more self-consistent results at higher mass-loss rates.
This is expected, since the mass-loss rate—derived from the optical depth—depends strongly on the IR excess of the stellar flux. 
When the SED is normalized, sensitivity to the absolute IR excess is reduced, while sensitivity to the spectral shape at shorter wavelengths increases. 
In contrast, the $T_{\rm eff}$ values are absolutely comparable.

\end{appendix}

\end{document}